\documentclass[english,prd,notitlepage,nofootinbib]{revtex4-1}
\usepackage[T1]{fontenc}
\usepackage[latin9]{inputenc}
\usepackage{babel}
\usepackage{amsmath}
\usepackage{amssymb}
\usepackage{graphicx}
\usepackage{esint}
\usepackage[unicode=true,pdfusetitle,
 bookmarks=true,bookmarksnumbered=false,bookmarksopen=false,
 breaklinks=false,pdfborder={0 0 1},backref=false,colorlinks=false]
 {hyperref}

\makeatletter

\providecommand{\tabularnewline}{\\}

\@ifundefined{textcolor}{}
{%
 \definecolor{BLACK}{gray}{0}
 \definecolor{WHITE}{gray}{1}
 \definecolor{RED}{rgb}{1,0,0}
 \definecolor{GREEN}{rgb}{0,1,0}
 \definecolor{BLUE}{rgb}{0,0,1}
 \definecolor{CYAN}{cmyk}{1,0,0,0}
 \definecolor{MAGENTA}{cmyk}{0,1,0,0}
 \definecolor{YELLOW}{cmyk}{0,0,1,0}
}

\usepackage{babel}
\usepackage{epsfig}
\usepackage{epsf}

\providecommand{\tabularnewline}{\\}

\@ifundefined{textcolor}{}{%
 \definecolor{BLACK}{gray}{0}
 \definecolor{WHITE}{gray}{1}
 \definecolor{RED}{rgb}{1,0,0}
 \definecolor{GREEN}{rgb}{0,1,0}
 \definecolor{BLUE}{rgb}{0,0,1}
 \definecolor{CYAN}{cmyk}{1,0,0,0}
 \definecolor{MAGENTA}{cmyk}{0,1,0,0}
 \definecolor{YELLOW}{cmyk}{0,0,1,0}
}

\usepackage{babel}

\parskip=\itemsep               
\usepackage{colortbl}

\providecommand{\tabularnewline}{\\}

\@ifundefined{textcolor}{}{%
 \definecolor{BLACK}{gray}{0}
 \definecolor{WHITE}{gray}{1}
 \definecolor{RED}{rgb}{1,0,0}
 \definecolor{GREEN}{rgb}{0,1,0}
 \definecolor{BLUE}{rgb}{0,0,1}
 \definecolor{CYAN}{cmyk}{1,0,0,0}
 \definecolor{MAGENTA}{cmyk}{0,1,0,0}
 \definecolor{YELLOW}{cmyk}{0,0,1,0}
}
\definecolor{cyan}{RGB}{153,255,153}

\makeatother

\begin{document}

\title{BFKL pomeron with massive gluons and running coupling}

\author{Eugene Levin$^{a,b}$, Lev Lipatov$^{c,d}$ and Marat Siddikov$^{a}$ }

\address{$^{a}$Departamento de Física, Universidad Técnica Federico Santa
María, y Centro Científico - Tecnológico de Valparaíso, Casilla 110-V,
Valparaíso, Chile}

\affiliation{$^{b}$Department of Particle Physics, School of Physics and Astronomy,
Tel Aviv University, Tel Aviv, 69978, Israel }

\affiliation{$^{c}$Theoretical Physics Department, Petersburg Nuclear Physics
Institute, Orlova Roscha, Gatchina, 188300, St. Petersburg, Russia }

\affiliation{$^{d}$Physics Department, St.Petersburg State University, Ulyanovskaya
3, St.Petersburg 198504, Russia }

\preprint{USM-TH-XXX}


\begin{abstract}
In this paper we proceed with the study of the Pomeron spectrum, by
solving numerically the BFKL equation with massive gluons and running
coupling. The spectrum of Regge singularities is discrete and the
leading Pomeron has a considerable dependence on nonperturbative effects,
for which we use Higgs mechanism as a model. We cross-checked this
result with variational method and confirmed the infrared sensitivity
of leading Pomeron. This fact is related to the infrared instability
of the BFKL equation in QCD, with a running coupling. The subleading
poles have a mild sensitivity to the soft physics, and are well described
by known semiclassical methods. We also discuss the dependence on various
prescriptions of the running coupling arguments. 
\end{abstract}
\maketitle
\tableofcontents{}

\section{Introduction}

The BFKL Pomeron\cite{BFKL,LI} is a structural element of all effective
theories for high energy QCD. It is a solution to the BFKL evolution
equation which sums large log $\left(\bar{\alpha}_{S}\ln(1/x)\right)^{n}$
terms in the perturbative QCD approach, and gives the scattering amplitude
at high energies. This amplitude possesses two fundamental properties:
the power-like energy dependence of the scattering amplitude~%
\footnote{This behavior is valid for all sizes of interacting colorless dipoles.
We refer to the book of Ref.\cite{KLB} for discussion on why in QCD
the colorless dipoles are the correct degrees of freedom at high energy. %
}, $A\propto\left(1/x\right)^{\omega_{0}}$, where $\omega_{0}=4\ln2\bar{\alpha}_{S}$,
and the growth of sizes of the typical dipoles at high energy. The
former feature contradicts the Froissart theorem~\cite{FROI} , while
the second imposes constraint on the applicability of the perturbative
approach, when we approach the confinement region. Both problems have
been solved in CGC/saturation approach~%
\footnote{See book\cite{KLB} for the review%
}, though the large impact parameter ($b$) dependence of the scattering
amplitude still remains an open question.

As it has been discussed in Refs.\cite{KW1,KW2,KW3,FIIM}, the scattering
amplitude at fixed $b$ in this approach should satisfy the unitarity
constraint of being smaller than unity, but since the radius of interaction
increases as a power of energy, eventually this leads to violation
of the Froissart bound~\cite{FROI}. Such power-like behavior of
the radius is a direct consequence of the perturbative QCD approach,
and stems from large impact parameter behavior of the BFKL Pomeron\cite{BFKL,LI}.
Therefore, we have to find how the confinement of quarks and gluons,
will affect the large $b$ behavior of the scattering amplitude. Since
we are interested in the behavior of the scattering amplitude at large
$b$, where the amplitude is small, the saturation effects can be
neglected and one can introduce the non-perturbative corrections directly
to the BFKL kernel.

In general, the main problem which one needs to solve, is the influence
of the unknown infrared behavior on the BFKL Pomeron. It is known
that due to a running QCD coupling in a problem, there appears a dimensional
scale, $\Lambda_{QCD}$. According to previous studies in~\cite{GLR,LI,LEAS,LERUN,KLR1,KLR2,KLR3,KLR4,KLRW1,KLRW2,LEPO}
due to the running coupling constant, the spectrum of the problem
becomes discrete for positive $\omega$, with infinitely many Regge
poles with quantum numbers of Pomeron, but also the kernel has unbound
from below continuous spectrum for negative $\omega$.

As it has been shown by one of us~\cite{LI}, the spectrum of these
Regge poles depends on the behavior of the scattering amplitude in
the confinement region. Since the theory of confinement is still in
the development stage, nowadays there are two approaches: phenomenological
extraction of nonperturbative effects from the experimental data,
or their evaluation in an effective model with built-in confinement.
The first approach parametrizes the information on confinement in
terms of the nonperturbative infrared phases~\cite{KLRW1,KLRW2,LEPO},
and produces a reasonable description of HERA data on deep inelastic
scattering~\cite{KLRW1,KLRW2,LEPO}. However, this analysis is complicated
by the fact that due to a limited energy range and not sufficiently
small $\langle x\rangle\sim10^{-3}$ in HERA kinematics, a very large
number of poles in a series need to be included in the fit, affecting
the precision of the extracted parameters. In this paper we address
the problem using the second approach, namely studying the spectrum
of the BFKL Pomeron in the particular model of infrared behavior,
the non-abelian theory with the Higgs mechanism for the gluon mass
generation~\cite{LLS1,LLS2}. This gauge invariant and normalizable
model has a correct large-$b$ behavior $\propto\exp\left(-m\, b\right)$
of the scattering amplitude. At short distances $r\ll m$, where $m$
is the effective gluon mass, this model transforms smoothly into perturbative
QCD, while at distances $r\sim1/m$ the gluon acquires a finite mass,
in agreement with what was found for correlation functions from eliminating
Gribov's copies~\cite{GRCO} (see Refs.\cite{GRCTH,GRCREV,GRCMASSG}).
We wish to stress that a gauge theory with the Higgs mechanism leads
to a good description of the gluon propagator calculated in lattice
approach~ \cite{GRCLAT}, with gluon mass $m=0.54\,{\rm GeV}$.

The first attempt to find the spectrum of the BFKL Pomeron Higgs model
with running $\bar{\alpha}_{S}$, was undertaken in~\cite{LICRO}
by one of us (see also \cite{KLR1}), in the semiclassical approach.
However, the underlying assumptions of the semiclassical approach
are not fulfilled for all momenta, and for the leading intercept ,
this uncertainty might be substantial. For this reason, in this paper
we cross-check the results and solve numerically the BFKL equation
with a running QCD coupling constant. We study how the infrared behavior
of a theory influences its spectrum. Finally, we study the spectrum
with the so-called triumvirate form of the running $\alpha_{S}$ \cite{BRAS,LEAS,KOAS,BAAS}
used in~\cite{GLR,LI,LEAS,LERUN,KLR1,KLR2,KLR3,KLR4,KLRW1,KLRW2,LEPO},
and demonstrate that it has qualitatively the same spectrum. We expect
that similar results are valid for all other forms of running couplings,
for example~in BLM approach~\cite{Brodsky:1998kn,Brodsky:2002ka}.

The paper is structured as follows. In Section~\ref{sec:Theory}
we briefly recapitulate the main properties of the BFKL equation with
running coupling, as well as analyze different schemes of infrared
regularization. In Section~\ref{sec:Results} we present our results,
and analyze their dependence on the choice of the regularization scheme,
as well as on the low-energy confinement model. We use three different
approaches: the semi-classical approximation (based on the method
of steepest descent), the variational method (based on exact solution
to the BFKL equation in diffusion approximation) and the numerical
study of the spectrum in the lattice. Finally, in Section~\ref{sec:Conclusions}
we draw conclusions.

\section{Theoretical framework}

\label{sec:Theory}

\subsection{BFKL equation and running coupling $\bar{\alpha}_{S}\left(k,\, k'\right)$}

\label{sub:BFKL}

The BFKL equation with running $\bar{\alpha}_{S}$ can be written
as an eigenvalue problem

\begin{equation}
\omega\Phi_{\omega}(k)\,\,=\,\,-\,{\cal {\hat{H}}}\Phi_{\omega}\,\,=\,\,\int\frac{d^{2}k'}{2\pi}\bar{\alpha}_{S}\left(k,k'\right)K\left(k,k'\right)\Phi_{\omega}\left(k'\right),\label{GRA1}
\end{equation}

where the rapidity ($Y$) dependent scattering amplitude $\mathcal{A}\left(Y,k\right)$,
is related to $\phi_{\omega}$ as

\begin{equation}
\mathcal{A}\left(Y,k\right)\,\,=\,\,\int_{\epsilon-i\infty}^{\epsilon+i\infty}\frac{d\omega}{2\pi i}e^{\omega Y}\,\phi_{\omega}\left(k\right),\label{GRA2}
\end{equation}

and the kernel $K$ is defined as \cite{BFKL,LI}

\begin{equation}
\int d^{2}k'\,\bar{\alpha}_{S}\left(k,k'\right)K\left(k,k'\right)\phi_{\omega}\left(k'\right)\,\,=\int d^{2}k'\,\bar{\alpha}_{S}\left(k,k'\right)\left[\frac{\phi_{\omega}\left(k'\right)}{\left(\vec{k}-\vec{k}^{\,'}\right)^{2}}\,\,-\,\,\frac{k^{2}\,\phi_{\omega}\left(k\right)}{\left(\vec{k}-\vec{k}^{\,'}\right)^{2}\,\left(k^{2}\,+\,\left(\vec{k}-\vec{k}^{\,'}\right)^{2}\right)}\right].\label{K}
\end{equation}

For $\bar{\alpha}_{S}\left(k\right)=$ const, the conformal symmetry
of the kernel~(\ref{K}) allows us to find the eigenfunction of the
BFKL equation, in the following form

\begin{equation}
\phi_{\omega}\left(k\right)\,\,=\,\,\left(k^{2}\right)^{\gamma-1}\,\,\equiv\,\, e^{\left(\gamma-1\right)\, t}\,\,\equiv\,\,\left(k^{2}\right)^{-\frac{1}{2}\pm i\nu}\,\,\equiv\,\, e^{\left(-\frac{1}{2}\pm i\nu\right)\, t},\label{EIGF}
\end{equation}

where $t\,\,=\,\,\ln\left(k^{2}/\Lambda_{QCD}^{2}\right)$ and $\gamma=\frac{1}{2}\pm i\nu$
is a continuous parameter. It is related to eigenvalues as

\begin{equation}
\omega\left(\nu\right)\,\,=\,\,\bar{\alpha}_{S}\,\chi\left(\nu\right)\,\,=\,\,\bar{\alpha}_{S}\left(2\psi\left(1\right)\,-\,\psi\left(\frac{1}{2}+i\nu\right)\,-\,\psi\left(\frac{1}{2}-i\nu\right)\right),\label{EIGV}
\end{equation}

where $\psi(z)$ is the digamma function. The choice of the argument
of the running coupling is ambiguous~\cite{LI,LERUN,KLR1,KLR2,KLR3,KLR4,KLRW1,KLRW2,LICRO}
and is usually done absorbing parts of logarithmic NLO corrections
into a definition of $\alpha_{s}$. The simplest form suggested in~\cite{LICRO}
is 
\begin{equation}
\bar{\alpha}_{S}\left(k,\, k'\right)=\bar{\alpha}_{S}^{{\rm LO}}\left(k\right),\label{ALHOUT}
\end{equation}
where $\bar{\alpha}_{{\rm LO}}$ is the leading order running coupling
in pQCD. The virtue of~(\ref{ALHOUT}) is that it allows one to evaluate
the spectrum and wave functions using approximate semiclassical methods.
Although in~\cite{LICRO} it was considered for a fixed flavor number
scheme, in this paper we use a realistic leading order coupling with
a variable flavor number scheme~\cite{DeRujula:1976edq}

\begin{align}
\bar{\alpha}_{{\rm LO}}\left(k\right)\,\, & =\frac{1}{\beta_{0}^{{\rm light}}\,\ln\left(k^{2}/\Lambda_{{\rm QCD}}^{2}\right)-\frac{1}{6}\sum_{i}F\left(k,\, m_{i}\right)},\label{BAS}\\
\beta_{0}^{{\rm light}} & =\frac{11\, N_{c}-2\, N_{f}^{{\rm light}}}{12},
\end{align}

where $N_{c}$ is the number of colors, $N_{f}^{{\rm light}}$ is
the number of light quark flavors, a value of $\Lambda_{QCD}\approx150\,{\rm MeV}$
was fixed from $\alpha_{s}(M_{Z})\approx0.118$. The sum over $i$
in denominator of~(\ref{BAS}) runs over heavy flavors $c,\, b,\, t$,
and the threshold function $F$ is given by~\cite{DeRujula:1976edq}
\begin{equation}
F\left(k,\, m_{i}\right)\approx\ln\left(\frac{k^{2}+5m_{i}^{2}}{\Lambda_{{\rm QCD}}^{2}+5m_{i}^{2}}\right).
\end{equation}

As was shown in~\cite{FADLI}, the form~(\ref{ALHOUT}) allows one
to rewrite the NLO corrections to the BFKL equation, in a form similar
to (\ref{K}), provided we replace the kernel $K$ as 
\begin{eqnarray}
 &  & K\left(k,k'\right)\,\,=\,\,\bar{\alpha}_{S}\left(k\right)K^{\mbox{\tiny LO}}\left(k,k'\right)\,\,+\,\,\bar{\alpha}_{S}^{2}(k)\, K^{\mbox{\tiny NLO}}\left(k,k'\right),\label{GRA9}\\
 &  & \int d^{2}k'\, K^{\mbox{\tiny LO}}\left(k,k'\right)\,\left(k'^{2}\right)^{\gamma-1}\,=\,\chi\left(\gamma\right)\,\left(k^{2}\right)^{\gamma-1},\nonumber \\
 &  & \int d^{2}k'K^{\mbox{\tiny NLO}}\left(k,k'\right)\,\left(k'^{2}\right)^{\gamma-1}\,=\,\delta\left(\gamma\right)\,\left(k^{2}\right)^{\gamma-1},
\end{eqnarray}
where $\chi(\gamma)$ is defined in~(\ref{EIGV}), and an explicit
form of $\delta\left(\gamma\right)$ and $K^{\mbox{\tiny NLO}}\left(k,k'\right)$
may be found in~\cite{FADLI}. A symmetrized form of~Eq.~(\ref{GRA9})
was suggested in~\cite{FADLI}, 
\begin{equation}
\omega\phi_{\omega}(k)\,\,=\,\,\int\frac{d^{2}k'}{2\pi}\sqrt{\bar{\alpha}_{S}\left(k\right)\,\bar{\alpha}_{S}\left(k'\right)\,\,}K^{\mbox{\tiny LO}}\left(k,k'\right)\phi_{\omega}\left(k'\right),\label{GRA10}
\end{equation}
however, it can be reduced to~(\ref{ALHOUT}) by redefinition of
the wave function $\phi_{\omega}(k)\,=\,\sqrt{\bar{\alpha}_{S}(k)}\,\tilde{\phi}_{\omega}$.
The properties of this equation with 
\begin{equation}
\bar{\alpha}_{S}\left(k,\, k'\right)\,\,=\,\,\bar{\alpha}_{{\rm LO}}^{{\rm FFNS}}\left(k\right)\,=\,\frac{1}{\beta_{0}^{{\rm light}}\,\ln\left(k^{2}/\Lambda_{{\rm QCD}}\right)}\label{eq:alpha_QCD_simple}
\end{equation}

have been investigated in detail in~\cite{LI,LERUN,KLR1,KLR2,KLR3,KLR4,KLRW1,KLRW2}
using the semiclassical approximation. A more complicated form of
running coupling was suggested in~\cite{BRAS,LEAS,KOAS,BAAS} and
used in~\cite{GLR,LI,LEAS,LERUN,KLR1,KLR2,KLR3,KLR4,KLRW1,KLRW2,LEPO},

\begin{equation}
\bar{\alpha}_{S}\left(k,\, k'\right)\,\,=\,\,\frac{\bar{\alpha}_{{\rm LO}}\left(\vec{k}-\vec{k}^{\,'}\right)\,\bar{\alpha}_{{\rm LO}}\left(k'\right)}{\bar{\alpha}_{{\rm LO}}\left(k\right)}.\,\label{eq:alpha_QCD}
\end{equation}

The form of the QCD coupling given by~(\ref{eq:alpha_QCD}) is preferable,
compared to~(\ref{eq:alpha_QCD_simple}) because of the following
features: 
\begin{itemize}
\item \emph{Gluon reggeization}: It was proven that Eq. (\ref{eq:alpha_QCD}),
rewritten for the octet $t$-channel state and for momentum transferred
$q_{T}\not=0$, satisfies the bootstrap equation~\cite{BRAS,LEAS},
\emph{i.e.} leads to the gluon reggeization as expected on general
grounds~(see Refs.~\cite{KOAS,BAAS}). 
\item \emph{Summation of Feynman diagrams for large number of flavors $N_{f}\gg1$}:
A direct sum of the Feynman diagrams in the limit of large number
of fermions $N_{f}$ leads to the triumvirate structure both for octet
and for singlet exchanges in $t$-channel~\cite{KOAS,LEAS,BAAS}.
Due to renormalizability of QCD, this implies that the triumvirate
structure is preserved in the general case. 
\item \emph{Correspondence to NLO BFKL equation}: In Ref.~\cite{KOAS1}
it is shown that the triumvirate reproduces the term proportional
to~$\beta_{0}$ in the NLO BFKL kernel. 
\end{itemize}
After renormalization of the fields $\phi_{\omega}\left(k\right)\to\alpha_{s}\left(k\right)\phi_{\omega}\left(k\right)$,
the coupling (\ref{eq:alpha_QCD}) in front of the BFKL kernel effectively
reduces to a function of a difference $\bar{\alpha}_{{\rm LO}}\left(\vec{k}-\vec{k}^{\,'}\right).$
However, as we will see below in subsection~(\ref{sub:Higgs}), such
transformation does not work for the Regge trajectory~(\ref{eq:ReggeTrajectory})
with $m_{H}=0$.

\subsection{Analytic solutions for massless BFKL case}

A general analysis of spectrum with a coupling constant~(\ref{eq:alpha_QCD},\ref{BAS})
is quite complicated, so in the pioneering papers~\cite{LI,LEAS,LERUN,KLR1,KLR2,KLR3}
a BFKL spectrum was analyzed with $\bar{\alpha}_{S}\left(k,k'\right)$,
given by Eq.~(\ref{eq:alpha_QCD_simple}) in a simplified fixed-flavor
number scheme (FFNS). In this case we can solve Eq.~ (\ref{GRA1})
analytically in a Mellin space, making a transform for $\phi_{\omega}\left(k\right)$,

\begin{equation}
\phi_{\omega}\left(k\right)\,\,=\,\,\int_{-\infty}^{\infty}\frac{d\nu}{2\pi}\,\tilde{\phi}_{\omega}\left(\nu\right)\, e^{\left(-\frac{1}{2}\,+\, i\nu\right)\, t}.\label{MT}
\end{equation}

The Eq.~ (\ref{GRA1}) for the Mellin image~$\tilde{\phi}_{\omega}\left(\nu\right)$
becomes ~\cite{KLR3,GLR}

\begin{equation}
i\beta_{0}\omega\frac{d\tilde{\phi}_{\omega}\left(\nu\right)}{d\nu}=\chi\left(\nu\right)\tilde{\phi}_{\omega}\left(\nu\right)\label{GRA3}
\end{equation}

which has solutions

\begin{align}
\tilde{\phi}_{\omega}\left(\nu\right) & =e^{-\frac{i}{\beta_{0}\omega}\int_{0}^{\nu}\chi\left(\nu'\right)d\nu'}.\label{GRA4}
\end{align}
In general (\ref{GRA4}) is defined up to a real phase. The inverse
Mellin transform yields for the BFKL wave function

\begin{eqnarray}
\phi_{\omega}\left(t\right)\,\, & = & \,\,\int_{-\infty}^{\infty}\frac{d\nu}{2\pi}\, e^{-\frac{i}{\beta_{0}\omega}\int_{0}^{\nu}\chi\left(\nu'\right)d\nu'\,+\,\left(-\frac{1}{2}+i\nu\right)t}\,\label{GRA5}\\
 & = & \,\int_{-\infty}^{\infty}\frac{d\nu}{2\pi}\,\Bigg(\frac{\Gamma\left(\frac{1}{2}+i\nu\right)}{\Gamma\left(\frac{1}{2}-i\nu\right)}\, e^{-2i\psi(1)\nu}\Bigg)^{\frac{1}{\beta_{0}\,\omega}}\, e^{\left(-\frac{1}{2}+i\nu\right)t}=\,\,\int_{-\infty}^{\infty}\frac{d\nu}{2\pi}\, e^{\left(-\frac{1}{2}+i\nu\right)t\,+i\varphi^{\mbox{\tiny pert}}\left(\omega,\nu\right)},\nonumber \\
\varphi^{\mbox{\tiny pert}}\left(\omega,\nu\right) & \equiv & \frac{1}{\beta_{0}\,\omega}\left({\rm Arg}\left[\frac{\Gamma\left(\frac{1}{2}+i\nu\right)}{\Gamma\left(\frac{1}{2}-i\nu\right)}\right]-2\,\psi(1)\,\nu\right)\label{eq:phi}
\end{eqnarray}
The formal solution~(\ref{GRA5}) should be supplemented with boundary
conditions. The first boundary condition stems from the behavior of
the wave function at large $t$. As has been shown in~\cite{LI,GLR},
in this kinematic region the solution of the BFKL equation should
match the solution for the DGLAP evolution equation\cite{DGLAP} in
double log approximation. In terms of the wave function this means
that at positive $\omega$, values of $\nu$ should be real. The second
boundary condition comes from the non-perturbative QCD approach, which
gives the wave function with the phase $\varphi^{{\rm non-pert}}\left(\omega\right)$
at fixed value of $t=t_{0}$. This condition for positive $\omega$
results in quantization of the spectrum \cite{LI} from the equation
\begin{equation}
\varphi^{{\rm non-pert}}\left(\omega\right)\,=\,\varphi^{{\rm pert}}\left(\omega,\nu\right).\label{EQDS}
\end{equation}

The choice of the value of $t_{0}$ is dictated by a requirement that
running QCD coupling $\alpha_{s}$ should be small enough to justify
application of the leading order BFKL approach. Since we impose two
boundary conditions at $t\to\infty$ and at $t=t_{0}$, for positive
$\omega$ the spectrum should be quantized. For negative $\omega$,
we need not impose any conditions at $t\to\infty$, so the spectrum
remains continuous.

In~\cite{LERUN,KLR3,LICRO} it was found in a semiclassical approach
that for large root number $j$ the spectrum is given by 
\begin{equation}
\omega_{j}\,\approx\,\frac{0.4085}{j-\frac{1}{4}+\phi^{{\rm nonpert}}/\pi\,\,-\,\,\nu\, t_{0}/\pi},\qquad j\gg1.\label{eq:SemiclassicalSpectrum}
\end{equation}

However, as we will show below, due to limitations of semiclassical
approach the leading pole which controls the high-energy behavior
of amplitudes in this scheme has the largest uncertainty. Additionally,
there is an uncertainty which stems from the region of small momenta,
where perturbative couplings in~(\ref{eq:alpha_QCD},\ref{BAS})
become inapplicable due to an infrared pole. This requires some regularization
at a low scale.

\subsection{Higgs mechanism}

\label{sub:Higgs}

The BFKL kernel~(\ref{K}) was obtained in the regime of asymptotically
large $|k|\gg\Lambda_{{\rm QCD}}$. However it is known that in the
regime of small momenta, nonperturbative effects affect drastically
all the partons, generating nonperturbative masses~\cite{Cornwall:1981zr}
and affecting their interactions~\cite{Maris:2003vk}. The non-abelian
theory with Higgs mechanism of mass generation is a particular (and
a relatively simple) model, in which nonperturbative effects only
generate an effective mass and affect the QCD behavior at large distances
$r\sim1/m$, where $m\approx540$~MeV is the effective gluon mass~\cite{GRCO,GRCTH,GRCREV,GRCMASSG}.
As was found in~\cite{GRCLAT}, this model leads to a good description
of the gluon propagator, calculated in lattice approach. In Standard
Model~\cite{Bartels:2006kr} this infrared regularization naturally
arises in the limit of vanishing Weinberg angle $\theta_{W}=0$.

\begin{figure}[ht]
\begin{centering}
\includegraphics[width=14cm]{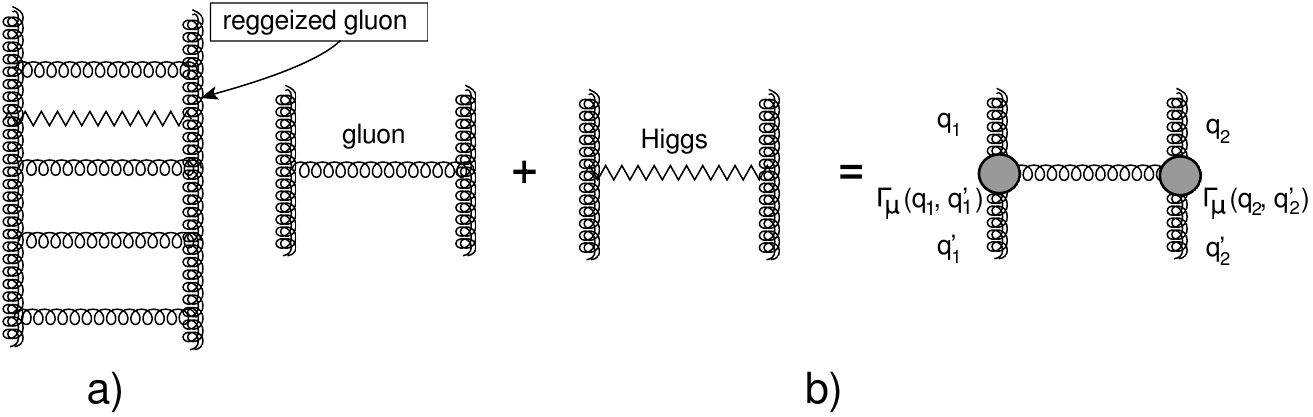} 
\par\end{centering}

\protect\protect\protect\caption{ The massive BFKL equation (\protect Fig.~ \ref{eq}-a)and its kernel
(\protect Fig.~ \ref{eq}-b)}

\label{eq} 
\end{figure}

The corresponding modification of a kernel (\ref{K}) in a massive
case takes the form (see \cite{LLS1,LLS2} and Fig.~\ref{eq})

\begin{eqnarray}
\omega\phi\left(k\right) & = & \frac{1}{\pi}\underbrace{\int d^{2}k'\,\bar{\alpha}_{S}\left(k,\, k'\right)\,\frac{\phi\left(k'\right)-\frac{1}{2}\frac{k^{2}+m_{H}^{2}}{k'^{2}+m_{H}^{2}}\phi\left(k\right)}{\left(k-k'\right)^{2}+m_{H}^{2}}}_{{\rm kinetic\, and\, potential\, energy\, terms}}\,\,-\,\,\underbrace{\frac{N_{c}^{2}+1}{2\pi N_{c}^{2}}\frac{m_{H}^{2}}{k^{2}+m_{H}^{2}}\int d^{2}k'\frac{\alpha_{{\rm LO}}^{2}\left(k'^{2}\right)}{\alpha_{{\rm LO}}\left(k^{2}\right)}\frac{\phi\left(k'^{2}\right)}{k'^{2}+m_{H}^{2}}}_{{\rm contact\, term}},\label{EQF-1}
\end{eqnarray}
where we denoted by $m_{H}$ the effective gluon mass. The coefficient
in front of $\phi(k)$ is sometimes referred to as a Regge trajectory
\begin{equation}
\omega\left(k\right)=-\frac{\left(k^{2}+m_{H}^{2}\right)}{2\pi}\int d^{2}k'\,\frac{\bar{\alpha}_{S}\left(k,\, k'\right)}{\left[\left(k-k'\right)^{2}+m_{H}^{2}\right]\left[k'^{2}+m_{H}^{2}\right]}.\label{eq:ReggeTrajectory}
\end{equation}

For the special case of~(\ref{ALHOUT}), the equation ~(\ref{EQF-1})
simplifies to 
\begin{eqnarray}
\omega\phi\left(k\right) & = & \frac{\bar{\alpha}_{S}\left(k\right)}{\pi}\Bigg\{\underbrace{\int d^{2}k'\,\frac{\phi\left(k'\right)-\frac{1}{2}\frac{k^{2}+m_{H}^{2}}{k'^{2}+m_{H}^{2}}\phi\left(k\right)}{\left(k-k'\right)^{2}+m_{H}^{2}}}_{{\rm kinetic\, and\, potential\, energy\, terms}}\,\,-\,\,\underbrace{\frac{N_{c}^{2}+1}{2\pi N_{c}^{2}}\frac{m_{H}^{2}}{k^{2}+m_{H}^{2}}\int d^{2}k'\,\frac{\phi\left(k'^{2}\right)}{k'^{2}+m_{H}^{2}}}_{{\rm contact\, term}}\Bigg\}.\label{EQF-2}
\end{eqnarray}
As we can see, both~(\ref{EQF-1}) and (\ref{EQF-2}) are not symmetric
w.r.t. an interchange of arguments $k$ and $k'$, which implies that
a kernel operator is no longer hermitian. However, its eigenvalues
remain real. Indeed, in the first term of (\ref{EQF-1},\ref{EQF-2})
we can symmetrize the kernel by a redefinition of the wave function
$\phi(\kappa)\to F(\kappa)\phi(\kappa)$, where $F(k)=\alpha(k)$
for (\ref{EQF-1}) with coupling~(\ref{eq:alpha_QCD}), and $F(k)=\sqrt{\alpha(k)}$
for (\ref{EQF-2}). The contribution of the last (contact) term in
(\ref{EQF-1},\ref{EQF-2}) does not affect the eigenvalues~\cite{LLS1,LLS2}.

The equation~(\ref{EQF-1}) has been studied in detail in~\cite{LLS1,LLS2}
for the case when $\alpha_{s}\left(k,k'\right)=$const. In this case
the integration over the azimuthal angle can be easily done, yielding
\begin{eqnarray}
\hspace{-0.5cm} &  & E\phi\left(\kappa\right)\,\,=\label{EQF}\\
\hspace{-0.5cm} &  & \,\,\underbrace{\frac{\kappa+1}{\sqrt{\kappa}\sqrt{\kappa+4}}\ln\frac{\sqrt{\kappa+4}+\sqrt{\kappa}}{\sqrt{\kappa+4}-\sqrt{\kappa}}\phi\left(\kappa\right)}_{{\rm kinetic\, energy\, term}}\,\,-\,\,\underbrace{\int_{0}^{\infty}\,\frac{d\kappa'\phi\left(\kappa'\right)}{\sqrt{(\kappa-\kappa')^{2}\,+\,2(\kappa+\kappa')+1}}}_{{\rm potential\, energy\, term}}\,\,+\,\,\underbrace{\frac{N_{c}^{2}+1}{2N_{c}^{2}}\frac{1}{\kappa+1}\int_{0}^{\infty}\frac{\phi\left(\kappa'\right)\, d\kappa'}{\kappa'+1}}_{{\rm contact\, term}},\nonumber 
\end{eqnarray}

where we introduced dimensionless variables and parameters

\begin{equation}
\kappa\,=\,\frac{k^{2}}{m_{H}^{2}};\,\,\,\,\,\,\,\,\,\,\,\,\,\,\kappa'\,=\,\frac{k'^{2}}{m_{H}^{2}};\,\,\,\,\,\,\,\,\,\,\,\,\, E\,=\,-\frac{\omega}{\bar{\alpha}_{S}};\,\,\,\,\,\,\,\,\,\,\,\,\,\bar{\alpha}_{S}\,=\,\frac{\alpha_{S}N_{c}}{\pi}.\label{VAR}
\end{equation}
It was found that the inclusion of a gluon mass does not affect the
eigenvalues of a problem, though changes the wave functions at small
momenta. Solving the problem on the lattice (see (\ref{sec:NumericalMethod})
for details), we found that for positive intercepts in a wide range
of $\kappa$, the wave functions may be approximated as

\begin{eqnarray}
\phi_{n}^{\mbox{ \tiny(approx)}}\left(\kappa\right)\,\, & \approx & \,\,\frac{{\rm const}}{\sqrt{\kappa+4}}\sin\Big(\beta\left(n\right)\, Ln\left(\kappa\right)+\phi(n)\Big),\label{APPWF}
\end{eqnarray}
where we introduced the notation 
\begin{equation}
Ln\left(\kappa\right)\,=\,\ln\Big(\frac{\sqrt{\kappa\,+\,4}\,\,+\,\,\sqrt{\kappa}}{\sqrt{\kappa\,+\,4}\,\,-\,\,\sqrt{\kappa}}\Big),
\end{equation}
and $\beta(n)$, $\varphi(n)$ are linear functions of a parameter
$n$, $\beta(n)=b\, n$, $\varphi(n)\approx b_{\phi}\,\beta(n)$.
The found value of a parameter $b_{\phi}$ does not depend on the
lattice choice and is given by $b_{\phi}\approx1.87$, but the parameter
$b$ depends on the lattice size, 
\begin{equation}
b=\frac{2.9}{\ln\left(k_{{\rm max}}^{2}/\left(k_{{\rm min}}^{2}+m_{H}^{2}\right)\right)}.\label{eq:b_def}
\end{equation}
In a continuum limit, a root number $n$ is proportional to a parameter
$\nu$ in~(\ref{EIGF},\ref{EIGV}), and~(\ref{APPWF}) takes the
form

\begin{equation}
\phi^{\mbox{ \tiny(approx)}}\left(\kappa,\,\nu\right)=\frac{\alpha\left(\nu\right)}{\sqrt{\kappa+4}}\sin\Big(\nu\, Ln\left(\kappa\right)+b_{\phi}\,\nu\Big).\label{APPWFCS}
\end{equation}
The approximation~(\ref{APPWFCS}) works for all $\nu$, for which
$\chi(\nu)$ remains positive. The solution~(\ref{APPWFCS}) is odd,
under $\nu\to-\nu$. Therefore, our eigenvalues are not degenerate
and we have one eigenfunction for each $\nu$.

\section{Results in a massive theory}

\label{sec:Results}

\subsection{Results with running coupling~(\ref{ALHOUT})}

\label{sub:LatticeResults} In this Section , we will analyze the
influence of the confinement region on the eigenvalues of the BFKL
equation with running $\bar{\alpha}_{s}$. In order to avoid an infrared
pole, we assume that at $t\,<\, t_{0}$ the coupling constant is frozen
(so the dynamics is described by (\ref{EQF})), whereas for $t\,\text{\ensuremath{\gg}}\, t_{0}$
we expect that the theory should reacquire the solution~(\ref{GRA5}).
Below we compare results for spectra found with two methods, a semiclassical
consideration and a lattice result.

\subsubsection{Semiclassical approximation}

\label{sec:QuasiClassics}

\begin{figure}[ht]
\begin{centering}
\includegraphics[width=12cm]{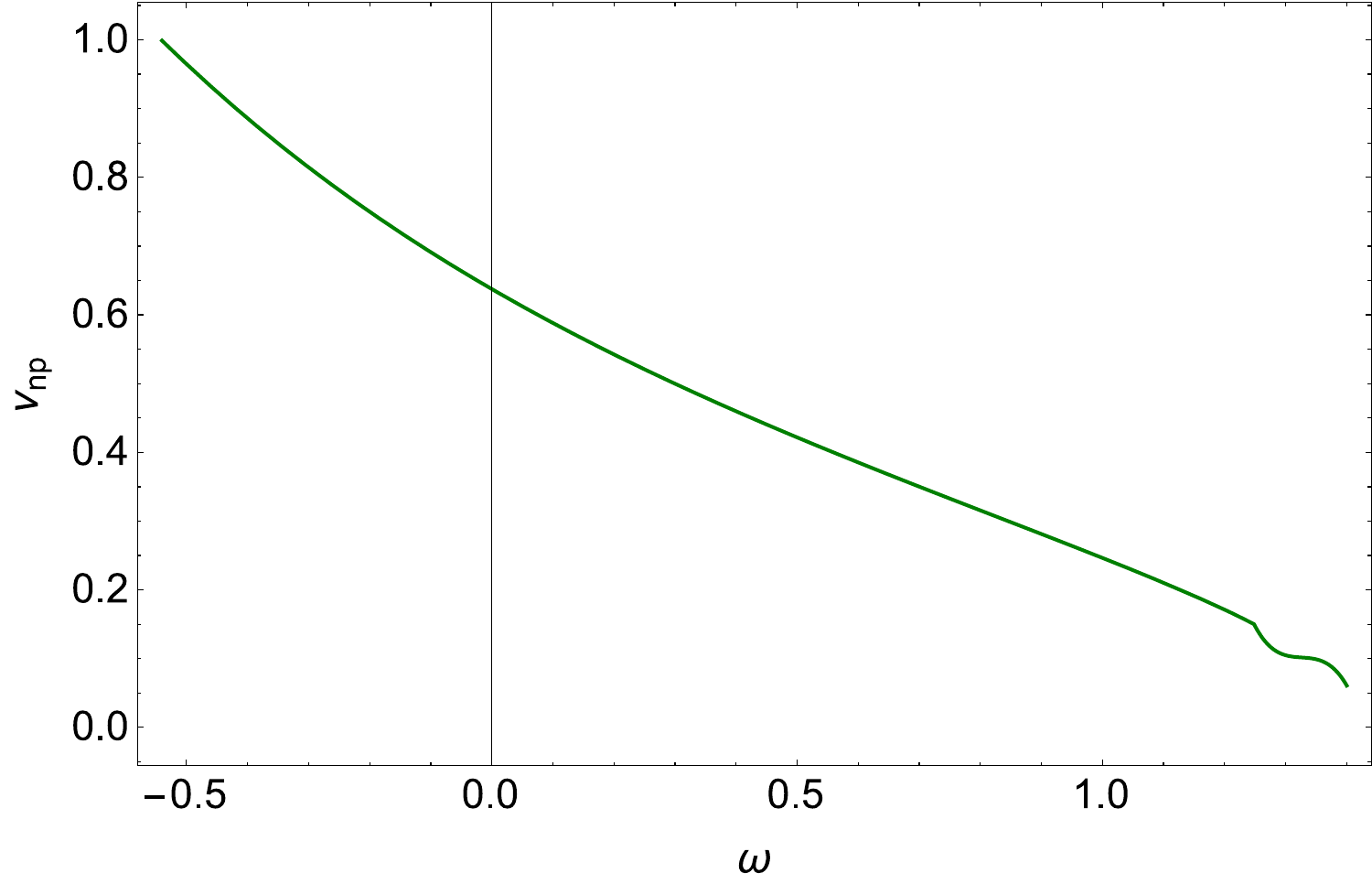} 
\par\end{centering}

\protect\protect\protect\caption{ $\nu_{{\rm np}}$ versus $\omega$ }

\label{nunp} 
\end{figure}

As was discussed in Section~\ref{sub:BFKL}, the analytic solution
of~(\ref{GRA1}) requires nonperturbative phase fixing~(\ref{EQDS}).
Due to the different character of wave functions and spectra (discrete
vs continua), the phase matching condition~(\ref{EQDS}) should be
supplemented with an additional requirement of equality of eigenvalues
\begin{equation}
\bar{\alpha}_{S}\left(m_{H}^{2}\right)\,\chi\left(\nu_{{\rm np}}\right)\,\,=\,\,\omega,\label{NUNP}
\end{equation}
where we use a notation $\nu_{{\rm np}}$ for the parameter which
appears in the non-perturbative regime. The dependence of $\nu_{{\rm np}}$
on $\omega$ which follows from~(\ref{NUNP}) is shown in Fig.~\ref{nunp}.

\begin{figure}[ht]
\begin{centering}
\includegraphics[width=10cm]{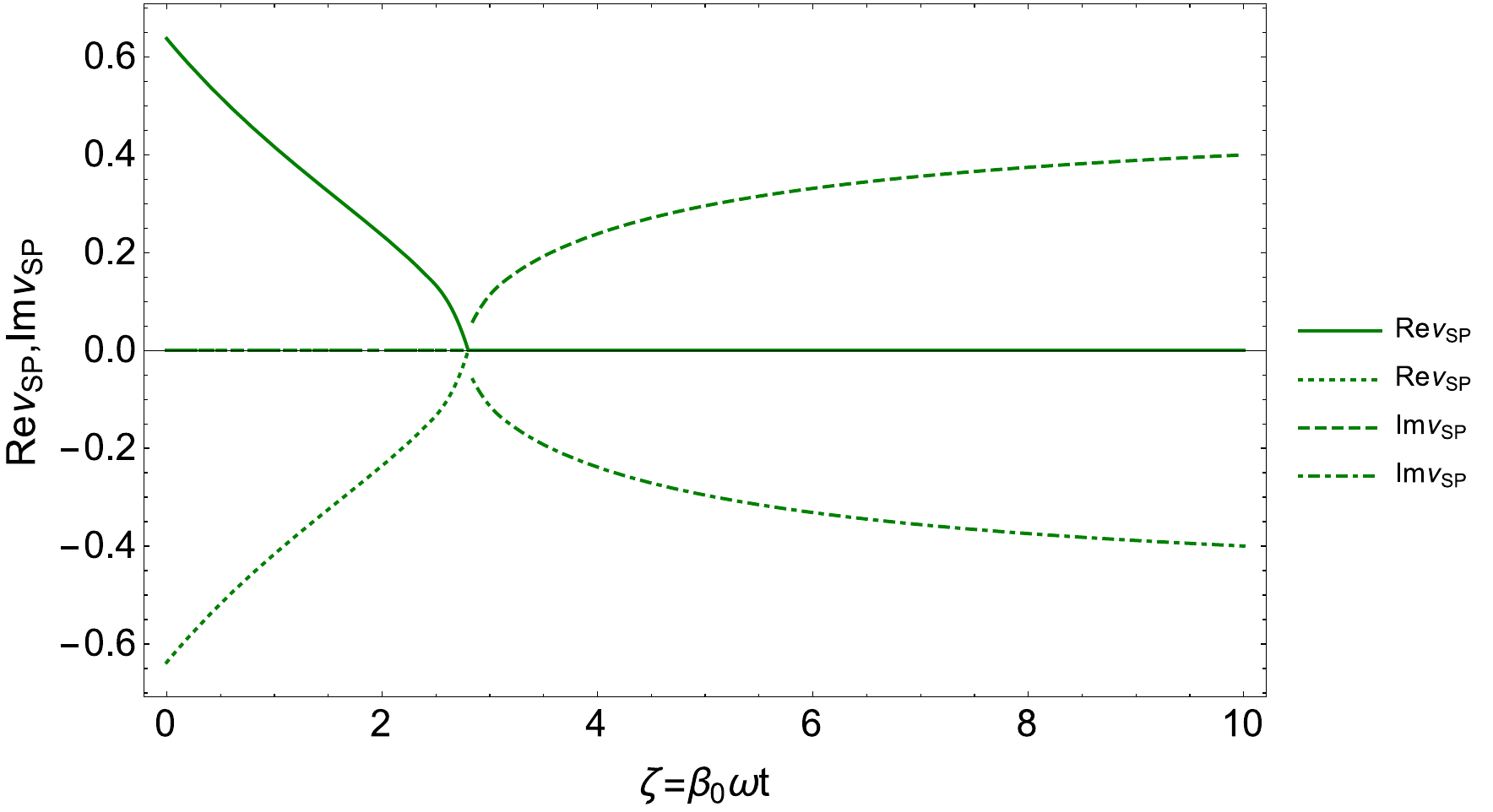} 
\par\end{centering}

\protect\protect\protect\caption{ $\mbox{Re}\nu_{\mbox{\tiny SP}}$ and $\mbox{Im}\nu_{\mbox{\tiny SP}}$
versus $\xi=\beta_{0}\omega t$ for $\omega\,>\,0$. At $\xi\leq\xi_{crit}$
$\nu_{\mbox{\tiny SP}}$ is real while for $\xi\geq\xi_{crit}$ is
pure imagine.}

\label{sp} 
\end{figure}

In our approach, we fix the nonperturbative phase from a solution~(\ref{APPWF}).
The choice of the matching point $t_{0}$ is arbitrary, provided the
running coupling $\alpha_{s}$ is small enough. Following the notation
introduced in previous papers~\cite{LERUN,KLR1,KLR2,KLR3}, we use
a variable $t=\ln\left(k^{2}/\Lambda_{QCD}^{2}\right)$. To evaluate
the integral over $\nu$ in~ (\ref{GRA5}) , we use the method of
steepest descent, as was suggested in~\cite{LERUN,KLR1,KLR2,KLR3}.
Since the coefficient in front of the $\mathcal{O}\left(\left(\nu-\nu_{SP}\right)^{2}\right)$
term in the expansion of $\varphi^{\mbox{\tiny pert}}\left(\omega,\nu\right)$
is imaginary for small-$t$, a contour should be taken parallel to
the lines bisecting the first and the second quadrants in the complex
plane. The saddle point is found from

\begin{equation}
\xi\,\,=\,\,\chi\left(\nu_{\mbox{\tiny SP}}\right),\label{SP}
\end{equation}

where $\xi\,=\,\beta_{0}\omega t$, and we are only interested in
roots closest to the real axis. As can be seen from definition of
$\chi(\nu)$ in~\ref{EIGV}, for real values of $\nu$ this function
is restricted from above by value $\sim4\ln2$, which implies that
for sufficiently large $\xi$, the relevant saddle-point $\nu_{\mbox{\tiny SP}}$
lies on the imaginary axis~(see Fig.~\ref{sp}). In this regime
instead of oscillations we have a homogenous decrease with $t$, and
a solution corresponds to a DGLAP solution~\cite{DGLAP} in the double
log approximation. As we will see below, the lattice wave functions
shown in the Figure~\ref{fig:R_WF}, at sufficiently large $t$ confirm
this pattern of behavior.

The phase matching condition~(\ref{EQDS}) in explicit form yields

\begin{equation}
\varphi^{\mbox{\tiny non-pert}}\left(\omega\right)+\pi j\,=\frac{\pi}{4}\,+\,\nu_{SP}\, t_{0}-2\frac{\psi\left(1\right)}{\beta_{0}\omega}\,\nu_{SP}\,+\,\frac{i}{\beta_{0}\omega}\ln\Bigg(\frac{\Gamma\left(\frac{1}{2}+i\nu_{SP}\right)}{\Gamma\left(\frac{1}{2}-i\nu_{SP}\right)}\Bigg),\quad j\in\mathbb{N}.\label{GRA6}
\end{equation}

The set of Eq.~ (\ref{SP},\ref{GRA6}) for positive $\omega$ leads
to quantization of spectrum of $\omega$ , whereas for negative $\omega$
the spectrum is continuous. Resolving Eq.~ (\ref{GRA6}) with respect
to $\omega$, we may obtain

\begin{equation}
\omega_{j}=\frac{1}{\beta_{0}}\frac{\Bigg\{-\,2\psi\left(1\right)\,\nu_{SP}\,+\,{\rm Im}\ln\Bigg(\frac{\Gamma\left(\frac{1}{2}+i\nu_{SP}\right)}{\Gamma\left(\frac{1}{2}-i\nu_{SP}\right)}\Bigg)\Bigg\}}{\pi\left(j-\frac{1}{4}\right)\,+\,\,\nu_{{\rm np}}\left(\omega_{j}\right)\Big(\, Ln\left(\frac{\Lambda_{QCD}^{2}}{m_{H}^{2}}\, e^{t_{0}}\right)+b_{\phi}\Big)-\,\,\nu_{SP}\, t_{0}}.\label{SM21}
\end{equation}

\begin{figure}[ht]
\begin{centering}
\includegraphics[width=10cm]{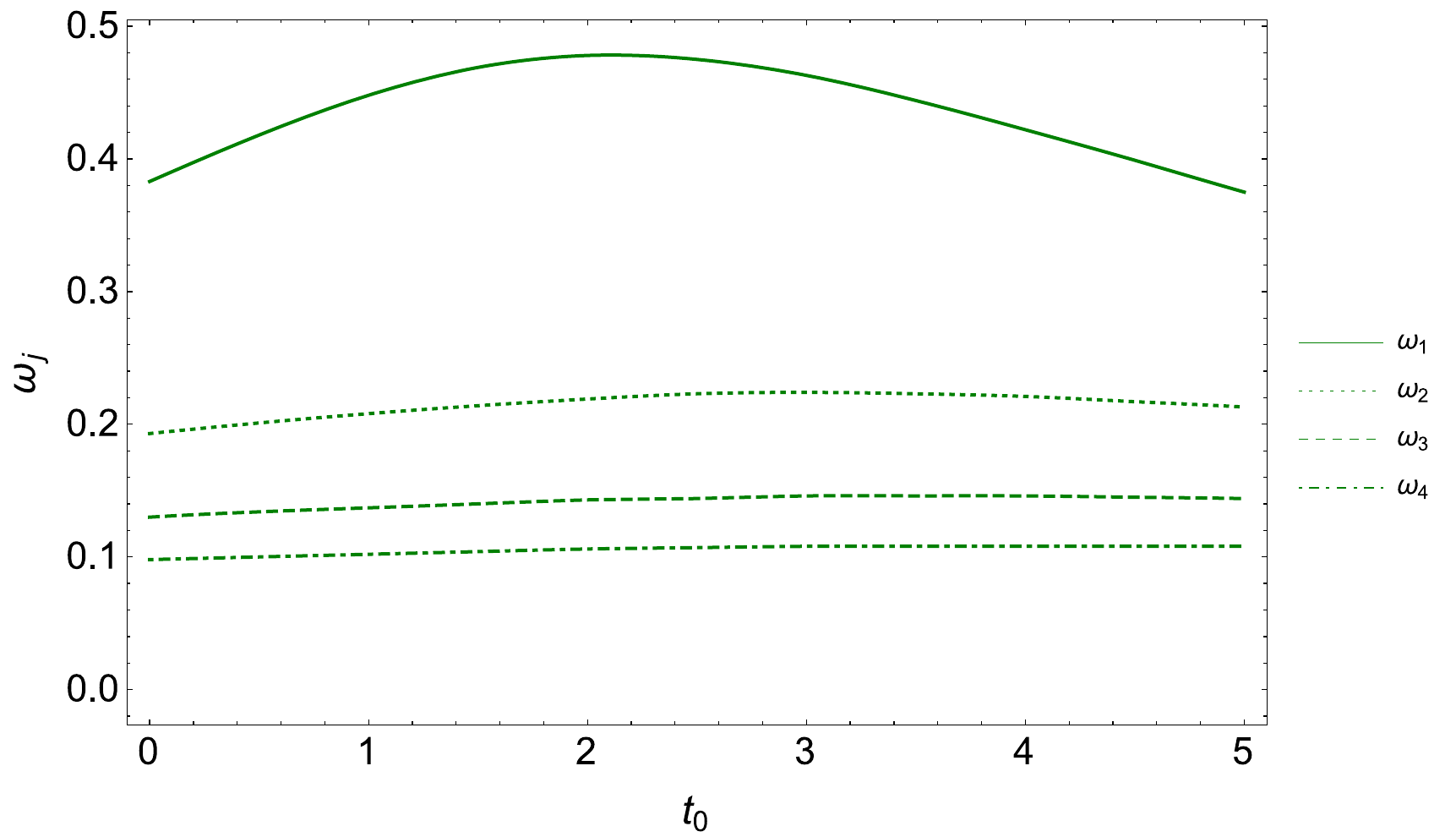} 
\par\end{centering}

\protect\protect\protect\caption{ The values of the first four eigenvalues of the BFKL equation in
semi-classical approach versus $t_{0}$. }

\label{om1} 
\end{figure}

The arbitrariness of the choice of matching point $t_{0}$ can be
reinterpreted in terms of uncertainty in the choice of a non-perturbative
phase, and would cancel if the $t$-dependence of $\varphi^{\mbox{\tiny pert}}$
and $\varphi^{\mbox{\tiny non-pert}}$ were the same. The dependence
of the eigenvalues on the choice of matching point $t_{0}$, is shown
in the Fig.~\ref{om1}. One can see that the sensitivity to the choice
of $t_{0}$ is very pronounced for the leading intercept, but decreases
rapidly for higher eigenvalues. Such a strong sensitivity to the infrared
dynamics (confinement region) can be understood from~(\ref{GRA5})~%
\footnote{See the Figure~\ref{fig:R_WF} below for the illustration of the
wave functions%
}: at asymptotically large momenta the wave functions decrease as $\propto1/k^{2}$,
while for constant $\bar{\alpha}_{S}$ the asymptotic behavior was
$\propto1/\sqrt{k^{2}}$. However, due to nodes for higher excited
states, the sensitivity to small momenta is diminished, while it remains
pronounced for the leading intercept. Similar sensitivity to the infrared
was obtained recently in~\cite{Ross:2016zwl}. In what follows for
the sake of definiteness we fix $t_{0}$ as $t_{0}=\ln\left(m_{H}^{2}/\Lambda_{{\rm QCD}}^{2}\right),$
which reduces~(\ref{SM21}) to

\begin{equation}
\omega_{j}=\frac{0.4085}{\left(j-\frac{1}{4}\right)\,+\,2.827\,\nu_{{\rm np}}\left(\omega_{j}\right)\,-\,1.65}.\label{SM4}
\end{equation}

This result agrees with~\cite{LICRO} for $t_{0}\approx0$.

The region of validity of the saddle-point approximation might be
estimated from the omitted $\mathcal{O}\left(\left(\nu-\nu_{SP}\right)^{3}\right)$
terms. For the saddle-point integral, the dominant contribution comes
from the region $|\delta\nu|<\delta\nu_{{\rm max}}$, where 
\begin{equation}
\delta\nu_{{\rm max}}=\sqrt{\beta_{0}\omega/\left(\frac{1}{2}\chi\left(\nu_{\mbox{\tiny SP}}\right)\right)}.
\end{equation}

A ratio $R$ of the $\mathcal{O}\left(\left(\nu-\nu_{SP}\right)^{3}\right)$-terms
to $\mathcal{O}\left(\left(\nu-\nu_{SP}\right)^{2}\right)$ -terms
in this region is given by

\begin{equation}
R=\omega/\left(\frac{1}{3!}\chi_{\nu}\left(\nu_{\mbox{\tiny SP}}\right)\frac{1}{\beta_{0}\omega}\left(\sqrt{\beta_{0}\omega/\left(\frac{1}{2}\chi\left(\nu_{\mbox{\tiny SP}}\right)\right)}\right)^{3}\right)^{2}.
\end{equation}

\begin{figure}[ht]
\begin{tabular}{ccc}
\includegraphics[width=7cm]{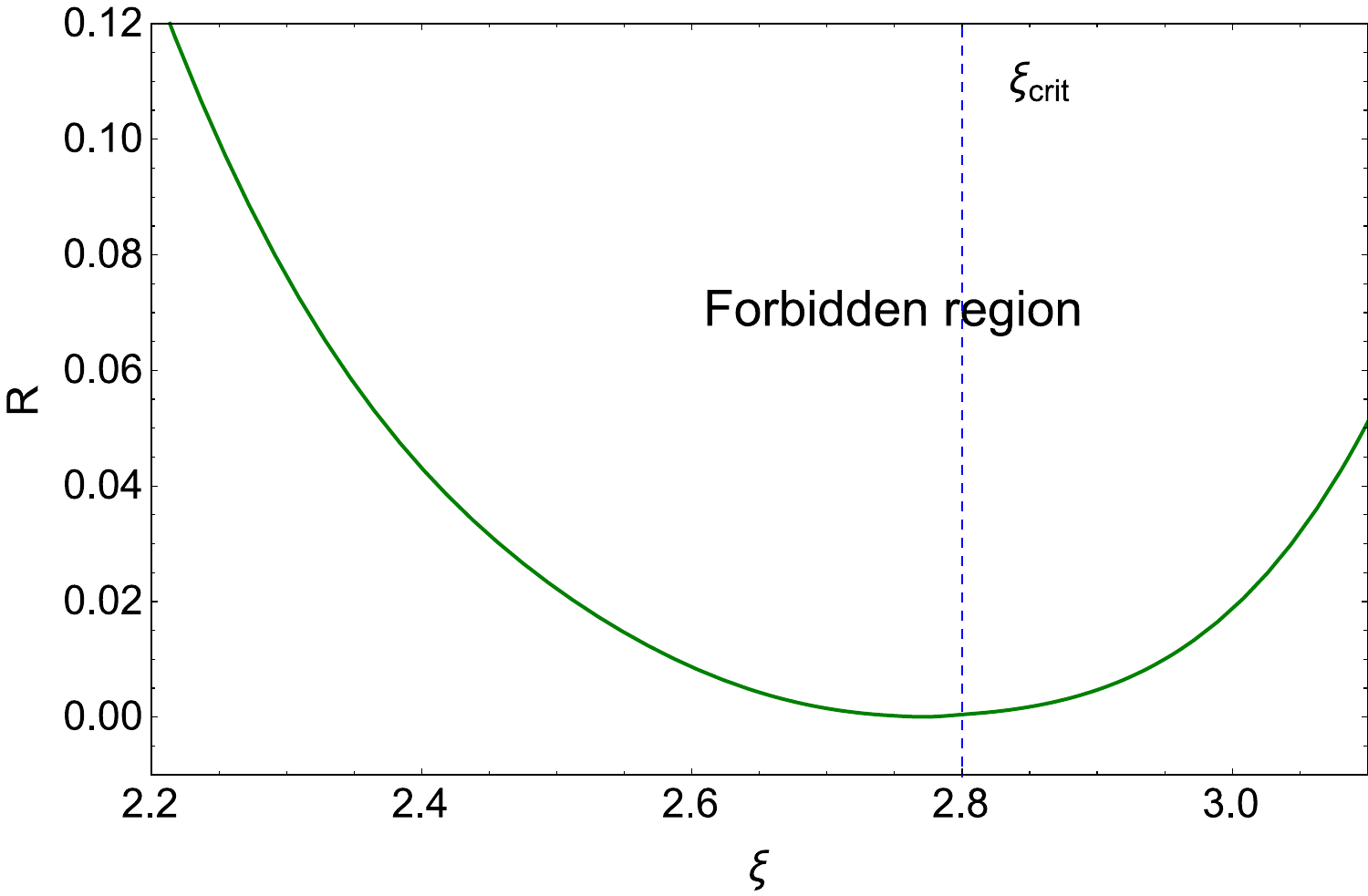}  & ~ ~ ~ ~ ~ ~ ~ ~ ~ ~ ~ ~ ~  & \includegraphics[width=7cm]{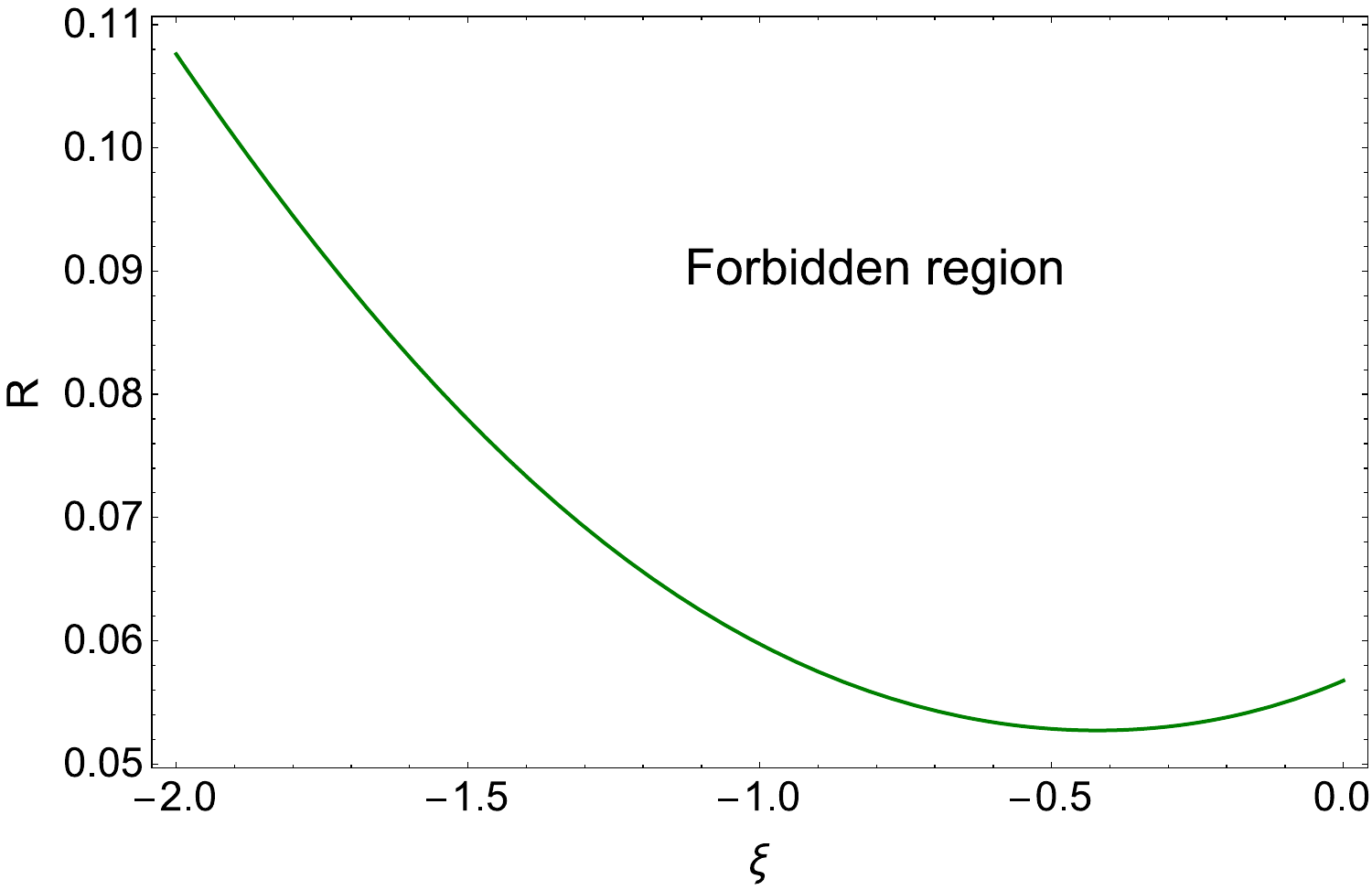}\tabularnewline
Fig.~ \ref{r}-a  &  & Fig.~ \ref{r}-b\tabularnewline
\end{tabular}

\protect\protect\protect\caption{ The values of $R$ versus $\xi$ for $\omega>0$(\protect Fig.~
\ref{r}-a)and for $\omega<0$ (\protect Fig.~ \ref{r}-b). For
$\omega\,<\, R$ we can trust the method of steepest descent. }

\label{r} 
\end{figure}

As we can see from the Fig.~\ref{r}, this ratio is small for $0<\omega\ll\omega_{{\rm crit}}$
and for negative $\omega$, but is not well-justified for the leading
eigenvalues, since these wave functions get large contributions from
a vicinity of $\xi_{{\rm crit}}$.

In the vicinity $\xi\approx\xi_{{\rm crit}}$ we can use the diffusion
approximation and expand $\chi(\nu)$ as 
\begin{equation}
\chi(\nu)\approx\chi_{0}\,\,-\,\, D_{0}\nu^{2}+\mathcal{O}\left(\nu^{4}\right)\label{eq:EIGV_expand}
\end{equation}
 with $\chi_{0}=4\ln2$ and $D_{0}=14\zeta(3)\approx16.828$~\cite{BFKL,KLB}.
Using this approximation, we can take the integral over $\nu'$ in
Eq.~(\ref{GRA5}) and express it in terms of the Airy functions (see
\cite{LERUN,KLR1,KLR2} for detailed discussions).

\subsubsection{Variational method}

\label{sec:VarMethod}In this section we discuss an approximate estimate
of the leading eigenvalues using the variational method. This method
reduces an integral equation~(\ref{EQF}) to a minimization problem
for the functional 
\begin{equation}
E[\phi]\,\,=\,\,\frac{\langle\phi|{\cal \hat{H}}|\phi\rangle}{\langle\phi|\phi\rangle},\label{VM1}
\end{equation}
where we use a variable $E=-\omega$. The variational method gives
an estimate from above for the energy (estimate from below for $\omega_{0}$),
with precision determined by the fact how close is the chosen trial
function to the true eigenfunction. The functional $\langle\phi|{\cal \hat{H}}|\phi\rangle$
which corresponds to (\ref{GRA1}) is given by 
\begin{equation}
\langle\phi|{\cal \hat{H}}|\phi\rangle\,\,=\,\,\int d^{2}k\,\phi(k)\,\bar{\alpha}_{S}(k)\,\int d^{2}k'\,\left[\frac{\phi_{\omega}\left(k'\right)}{\left(\vec{k}-\vec{k}^{\,'}\right)^{2}}\,\,-\,\,\frac{k^{2}\,\phi_{\omega}\left(k\right)}{\left(\vec{k}-\vec{k}^{\,'}\right)^{2}\,\left(k^{2}\,+\,\left(\vec{k}-\vec{k}^{\,'}\right)^{2}\right)}\right].\label{VM2}
\end{equation}
Inspired by the solution in the diffusion approximation~\cite{LEAS,LERUN,KLR2},
which is exact in the region of $\omega$ in the vicinity of leading
pole, we parametrize our trial function as 
\begin{equation}
t\,\geq\, t_{0}~~~~~~~\phi^{{\rm trial}}(k;\omega,a)\,\,=\,\, C\left(\omega\right)\,\Bigg(\frac{t}{e^{t}+a/\Lambda_{{\rm QCD}}^{2}}\Bigg)^{1/2}\,\mbox{Ai}\Bigg[\left(\frac{\omega}{D_{0}\beta_{0}}\right)^{1/3}\Big(\ln\left(e^{t}+a/\Lambda_{{\rm QCD}}^{2}\right)\,-\,\frac{\chi_{0}}{\omega}\Big)\Bigg],\label{VM3}
\end{equation}
where $\mbox{Ai}(x)$ is the Airy function, the parameters $\chi_{0}$
and $D_{0}$ are defined as Taylor expansion coefficients in~(\ref{eq:EIGV_expand}),
and for $t<t_{0}$, we use two models of boundary conditions

\begin{eqnarray}
 &  & \mbox{model I:}~~~~\phi^{{\rm trial}}(k;\omega,a)\,\,=\,\,0\,,\label{MOD1}\\
 &  & \mbox{model II:}~~~~~\phi^{{\rm trial}}(k;\omega,a)\,\,=\,\,\,\phi^{\mbox{ \tiny(approx)}}\left(\kappa,\,\nu_{{\rm np}}\right).\label{MOD2}
\end{eqnarray}
The function $\phi^{\mbox{ \tiny(approx)}}\left(\kappa,\,\nu_{{\rm np}}\right)$
in~(\ref{MOD2}) is given by equation (\ref{APPWFCS}) and $\nu_{{\rm np}}$
is determined by equation (\ref{NUNP}).

\begin{figure}[ht]
\begin{centering}
\includegraphics[width=14cm]{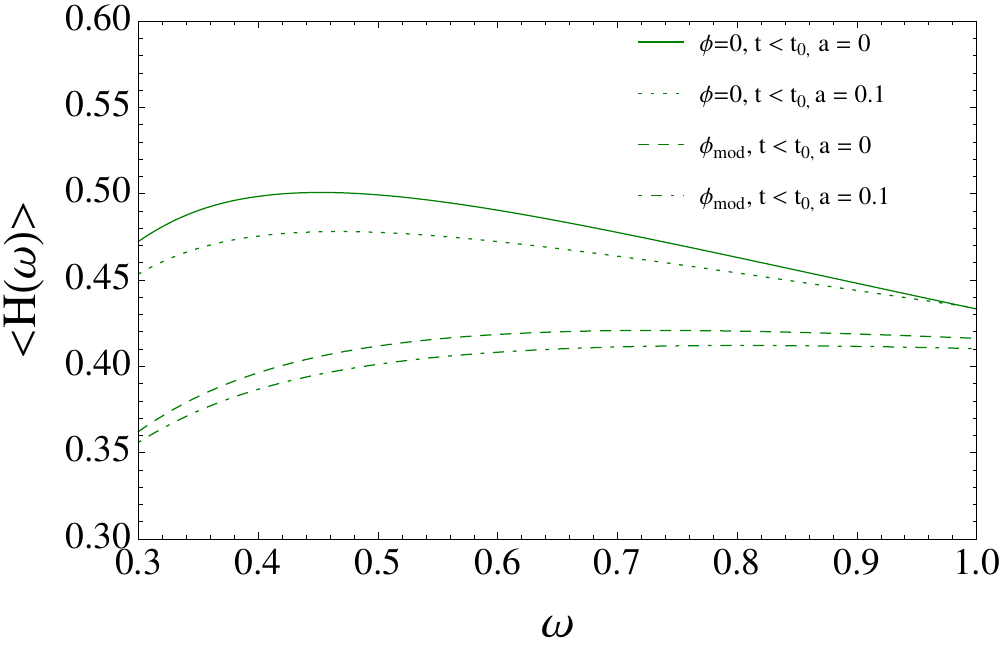} 
\par\end{centering}

\protect\protect\protect\caption{ Dependence of $E[\phi]$ given by equation ( \ref{VM1}) . $t_{0}=\ln(m_{H}^{2}/\Lambda_{{\rm QCD}})=2.5$.
$\langle H(\omega)\rangle$ denotes the functional of equation (\ref{VM1}).$\phi_{{\rm mod}}\equiv\phi^{\mbox{ \tiny(approx)}}$. }

\label{vm} 
\end{figure}

The model (\ref{MOD1}) assumes that confinement does not contribute
to the BFKL dynamics (so this region can be dropped completely) while
the model (\ref{MOD2}) corresponds to the confinement model discussed
in Section~\ref{sub:Higgs}. The former can be confronted with the
numerical solution of the next section (see columns with different
$t_{0}$ in the Table~\ref{tab:TenRoots}), while the latter we can
compare with the semi-classical approach. From Figure (\ref{vm})
we can see that we have maxima in both cases. For model~(\ref{MOD1})
the eigenvalue is equal to 0.5 for $a=0$, and differs from the lattice
result (column ``$t_{0}=2.5$'') by $(\omega_{{\rm exact}}-\omega_{{\rm var.~med.}})/\omega_{{\rm exact}}\,\approx\,16\%$.
The model II leads to $\omega_{0}=-E_{0}=0.42$ . Comparing this value
with the semiclassical approach we see $(\omega_{{\rm SCA}}-\omega_{{\rm var.~med.}})/\omega_{{\rm SCA}}\,\approx\,11-12\%$.
However, comparing with the exact solution given by column $\omega_{j}(k^{2}+m_{H}^{2})$
we see that $(\omega_{{\rm exact}}-\omega_{{\rm var.~med.}})/\omega_{{\rm exact}}\,\approx\,5\%$\emph{. }

Also, we analyze spectrum with more general trial function~(\ref{VM3})
given by 
\begin{equation}
\phi^{{\rm trial}}(k;\omega,a)\,\,=\,\,\mathcal{N}\,\Bigg(\frac{\ln\left(k^{2}/\Lambda_{{\rm QCD}}^{2}+a^{2}\right)}{k^{2}/\Lambda_{{\rm QCD}}^{2}+a^{2}}\Bigg)^{1/2}\,\mbox{Ai}\Bigg[b\,\Big(\ln\left(k^{2}/\Lambda_{{\rm QCD}}^{2}+a^{2}\right)\,-\, c\Big)\Bigg],\label{VM3-1}
\end{equation}
where $a,b,\, c$ are free parameters. After numerical minimization,
the best local minimum which we found corresponds to $a\approx1.02$,
$b\approx0.27$, $c\approx1.08$, and the leading intercept $\omega\approx0.58$
is only 2\% below the result found in the lattice. In a similar fashion
we may generalize the method and estimate subleading eigenvalues imposing
in~(\ref{VM1}) additional orthogonality conditions $\forall l<j,\quad\int d^{2}k\,\phi(k,\omega_{j})\phi(\omega_{l},k;)\,=\,0$,
however the precision of such extraction decreases with eigenvalue
number $j$ due to accumulation of errors of trial functions.

\subsubsection{Spectrum from the lattice}

\begin{figure}
\includegraphics[width=13cm]{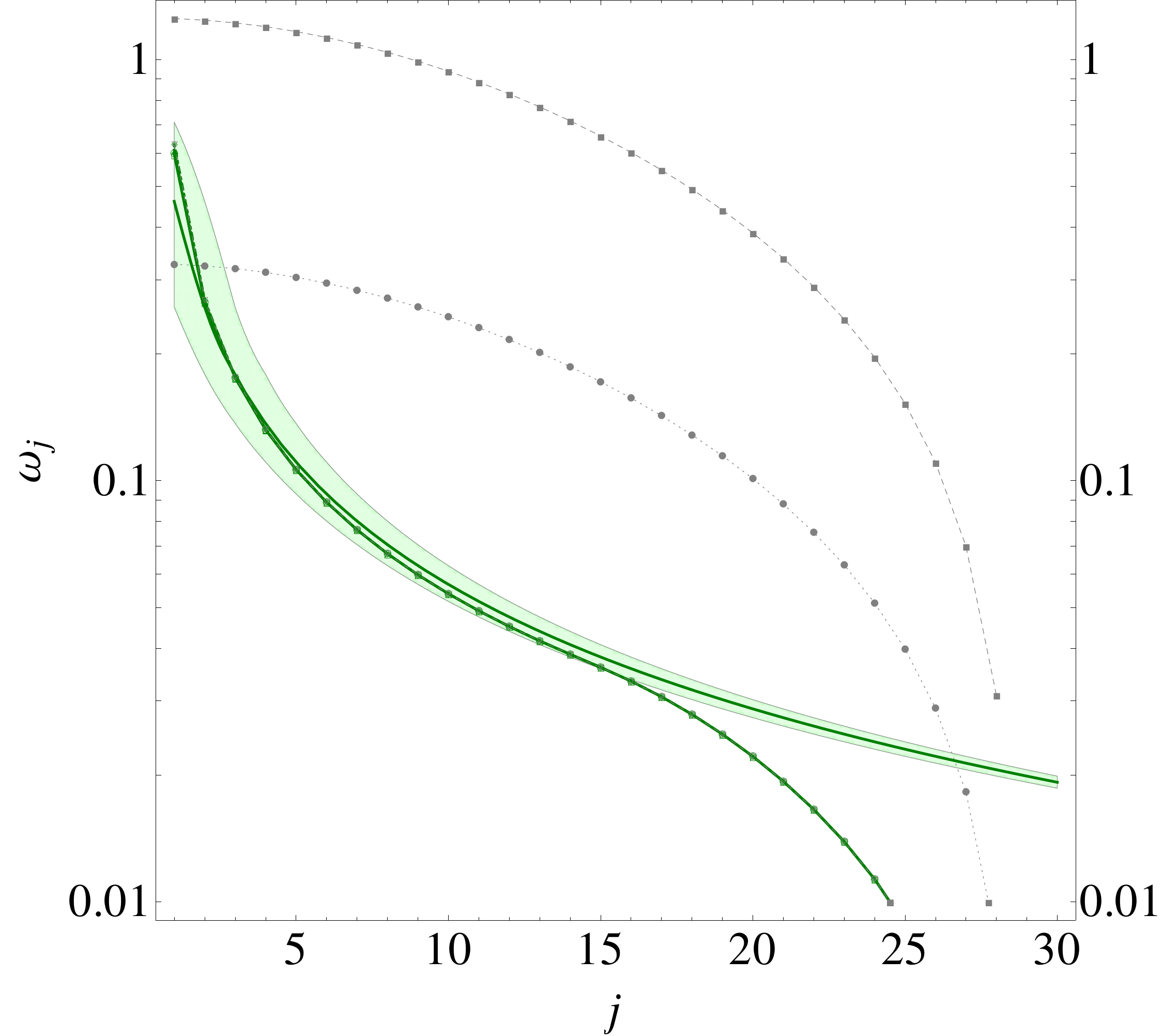}\\
 \includegraphics[width=13cm]{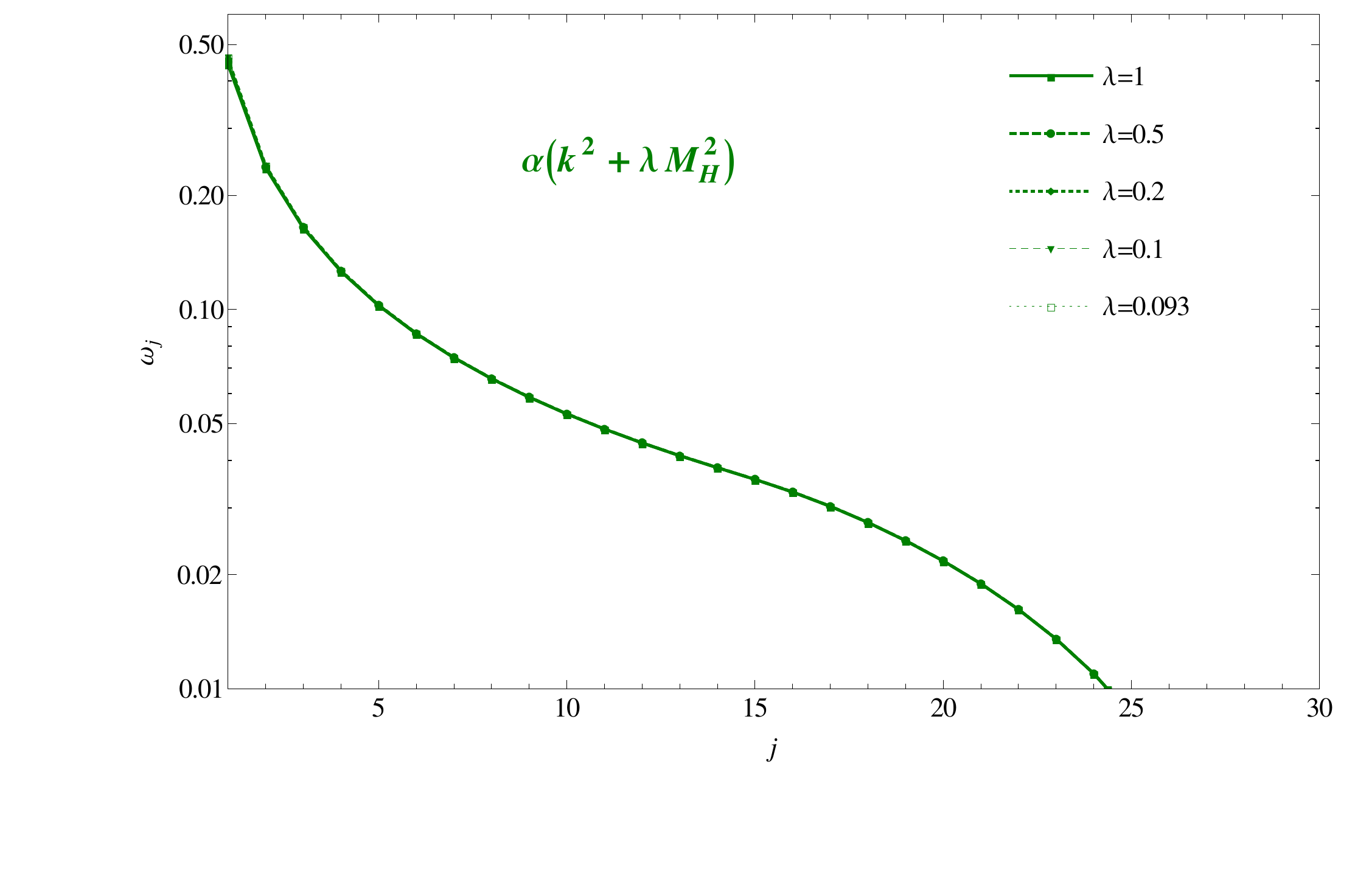}

\protect\protect\protect\caption{\label{fig:R_IR}(color online) Upper plot: Dependence of roots $\omega_{j}$
of BFKL on root number $j$ (variable $\nu$ in a continuum limit)
for different values of lower cutoff. Light green band corresponds
to semiclassical result with the central value evaluated according
to~(\ref{SM4}) and uncertainty of nonperturbative phase $\Delta\varphi=\pm\pi$
(see text for more details). The roots with different values of $t_{0}$
(dark green lines with different dashing patterns, corresponding to
$t_{0}$ between 0 and 2.5) cannot be distinguished from the plot;
numerical eigenvalues corresponding to different $t_{0}$ may be found
in Table~\ref{tab:TenRoots}. For reference we also put thin dashed
grey lines which correspond to evaluation with fixed coupling, taken
at scales $\alpha_{s}\left(m_{Z}\right)$ and $\alpha_{S}\left(m_{H}\right)$
respectively. Lower plot: Dependence of roots $\omega_{j}$ of BFKL
on root number $j$ (variable $\nu$ in a continuum limit) for different
schemes of freezing of a coupling constant. As we can see, all curves
almost coincide and are not discernible in the plot.}
\end{figure}

In order to cross-check our results, we also study the spectrum and
eigenfunctions using a lattice method introduced in~\cite{LLS1,LLS2}
(see Appendix~\ref{sec:NumericalMethod} for details). In this method
we no longer need to impose a nonperturbative phase fixing~(\ref{EQDS}),
and consider instead two alternative schemes of infrared regularization.
First, we assume that the gluon has no mass and set a minimal momentum
$\kappa_{min}=\Lambda_{{\rm QCD}}^{2}/m_{H}^{2}\exp\left(t_{0}\right)$.
As we can see from the second column of the Table~\ref{tab:TenRoots}
and upper panel of the Figure~(\ref{fig:R_IR}), the spectrum in
this case is discrete%
\footnote{In contrast to a discretized spectrum due to finite lattice size as
in Eq\@.(\ref{EQF}), the spectrum of~(\ref{EQF-2}) is truly discrete
because the distance between the neighboring eigenvalues is considerably
larger than in case of (\ref{EQF}), and does not decrease as a function
of upper lattice cutoff $\kappa_{{\rm max}}$ (see Section~\ref{sec:cutoff}
and Appendix~\ref{sec:NumericalMethod} for more details).%
} and has a very weak dependence on $t_{0}$. As expected, for all
$t_{0}$ the spectrum lies within uncertainty due to a choice of the
nonperturbative phase (light green bands), but the difference with
the semiclassical approximation (solid green central line) for the
lowest excited states, turns out to be about 20-30\% for the first
ten eigenvalues~(see columns one and five in Table~\ref{tab:TenRoots}).
An alternative regularization scheme is to consider~(\ref{EQF-2})
in a full range of momenta $k$, freezing the coupling constant in
the infrared region by adding some small scale $\lambda m_{H}^{2}$
to its argument. From the lower panel of Figure~(\ref{fig:R_IR})
we can see that the sensitivity to the choice of the scale is also
very weak , and agrees with eigenvalues shown in the upper panel.
The agreement with the semiclassical approach is quite good, and for
the first two eigenvalues, the two approaches agree within 6-7 percent.
However, the difference grows for smaller eigenvalues, and reaches
30 percent for the tenth eigenvalue. For the sake of comparison in
the same Figure~(\ref{fig:R_IR}) we also plotted eigenvalues of
the BFKL equation~(\ref{EQF}) with fixed coupling~$\bar{\alpha}_{s}$.
These curves illustrate, that the distance between the eigenvalues
due to lattice discretization of continuous spectrum is considerably
smaller, and thus the physical spectrum is discrete (see~Appendix~\ref{sec:NumericalMethod}
for more details).

\begin{table}
\setlength{\tabcolsep}{0.2pt}

\begin{tabular}{cccccccccccc}
\hline 
\rowcolor{cyan} $j$  & SCA  & \multicolumn{4}{c}{\cellcolor{cyan}Fig~(\ref{fig:R_IR})} & $\omega_{j}\left[\bar{\alpha}_{S}\left(k^{2}+m_{H}^{2}\right)\right]$  & \multicolumn{2}{c}{\cellcolor{cyan}Triumvirate~(\ref{eq:alpha_triumvirate})} & \multicolumn{3}{c}{\cellcolor{cyan}Triumvirate~(\ref{eq:alpha_eff_reg})}\tabularnewline
 & $\quad t_{0}$=2.5  & $\quad t_{0}$=0.1  & $\quad t_{0}$=0.8  & $\quad t_{0}$=1.8  & $\quad t_{0}$=2.5  &  & $\quad\lambda=0.1$  & $\quad\lambda=1$  & $\quad\lambda=0.05$  & $\quad\lambda=0.2$  & $\quad\lambda=1$ \tabularnewline
\hline 
1  & 0.475  & 0.626  & 0.609  & 0.599  & 0.595  & 0.442  & 0.509  & 0.418  & 0.479  & 0.430  & 0.381\tabularnewline
2  & 0.223  & 0.268  & 0.266  & 0.266  & 0.265  & 0.236  & 0.276  & 0.254  & 0.258  & 0.240  & 0.221\tabularnewline
3  & 0.144  & 0.176  & 0.176  & 0.175  & 0.175  & 0.164  & 0.187  & 0.177  & 0.178  & 0.168  & 0.158\tabularnewline
4  & 0.107  & 0.132  & 0.132  & 0.132  & 0.132  & 0.126  & 0.141  & 0.136  & 0.135  & 0.130  & 0.123\tabularnewline
5  & 0.085  & 0.106  & 0.106  & 0.106  & 0.106  & 0.102  & 0.114  & 0.110  & 0.109  & 0.105  & 0.101\tabularnewline
6  & 0.070  & 0.089  & 0.089  & 0.089  & 0.089  & 0.086  & 0.095  & 0.093  & 0.092  & 0.089  & 0.085\tabularnewline
7  & 0.060  & 0.076  & 0.076  & 0.076  & 0.076  & 0.074  & 0.081  & 0.080  & 0.079  & 0.077  & 0.074\tabularnewline
8  & 0.052  & 0.067  & 0.067  & 0.067  & 0.067  & 0.066  & 0.071  & 0.070  & 0.069  & 0.067  & 0.065\tabularnewline
9  & 0.046  & 0.060  & 0.060  & 0.060  & 0.060  & 0.059  & 0.063  & 0.062  & 0.061  & 0.060  & 0.058\tabularnewline
10  & 0.041  & 0.054  & 0.054  & 0.054  & 0.054  & 0.053  & 0.057  & 0.056  & 0.055  & 0.054  & 0.052\tabularnewline
\hline 
\end{tabular}\protect\protect\caption{\label{tab:TenRoots}The first ten eigenvalues $\omega_{j}$ of the
BFKL equation. First column (``SCA'') denotes results with the semiclassical
approximation (see \protect Eq.~(\ref{SM21})). The second group
of columns correspond to eigenvalues with different $t_{0}$ shown
in the upper pane of the Figure~(\ref{fig:R_IR}). The column marked
by $\omega_{j}\left[\bar{\alpha}_{S}\left(k^{2}+m_{H}^{2}\right)\right]$
is the solution to \protect Eq.~(\ref{EQF-1}) with running QCD
coupling frozen in low energy domain by shifting variable as $k^{2}\to k^{2}+M_{H}^{2}$.
The group of columns marked as ``Triumvirate'' corresponds to the
solution of~(\ref{EQF-1}) with running coupling~(\ref{eq:alpha_triumvirate})
with two different values of $\lambda$. Finally, the last group of
columns corresponds to triumvirate coupling regularized as~(\ref{eq:alpha_eff_reg}).
In all evaluations we used $M_{H}=540\,{\rm MeV}$ and $\Lambda_{QCD}=148\,{\rm MeV}$.}

\label{t1} 
\end{table}

To summarize, we conclude that results obtained with running coupling~~(\ref{ALHOUT})
are robust with respect to a change of the lower cutoff and freezing
scale. For the leading intercept, the sensitivity is largest, but
still remains within the phase uncertainty band of semiclassical approximation.
At large root number $j$, we observe a significant deviation of the
lattice results from semiclassical approximation. This can be understood
from~(\ref{eq:SemiclassicalSpectrum}): when the distance between
the roots becomes smaller than the energy discretization due to lattice
size, the method can no longer distinguish separate discrete roots
and jumps to negative branch (which has continuum spectrum). For this
reason for large root number $j$ we should trust the semiclassical
approximation.

\begin{figure}
\includegraphics[scale=0.75]{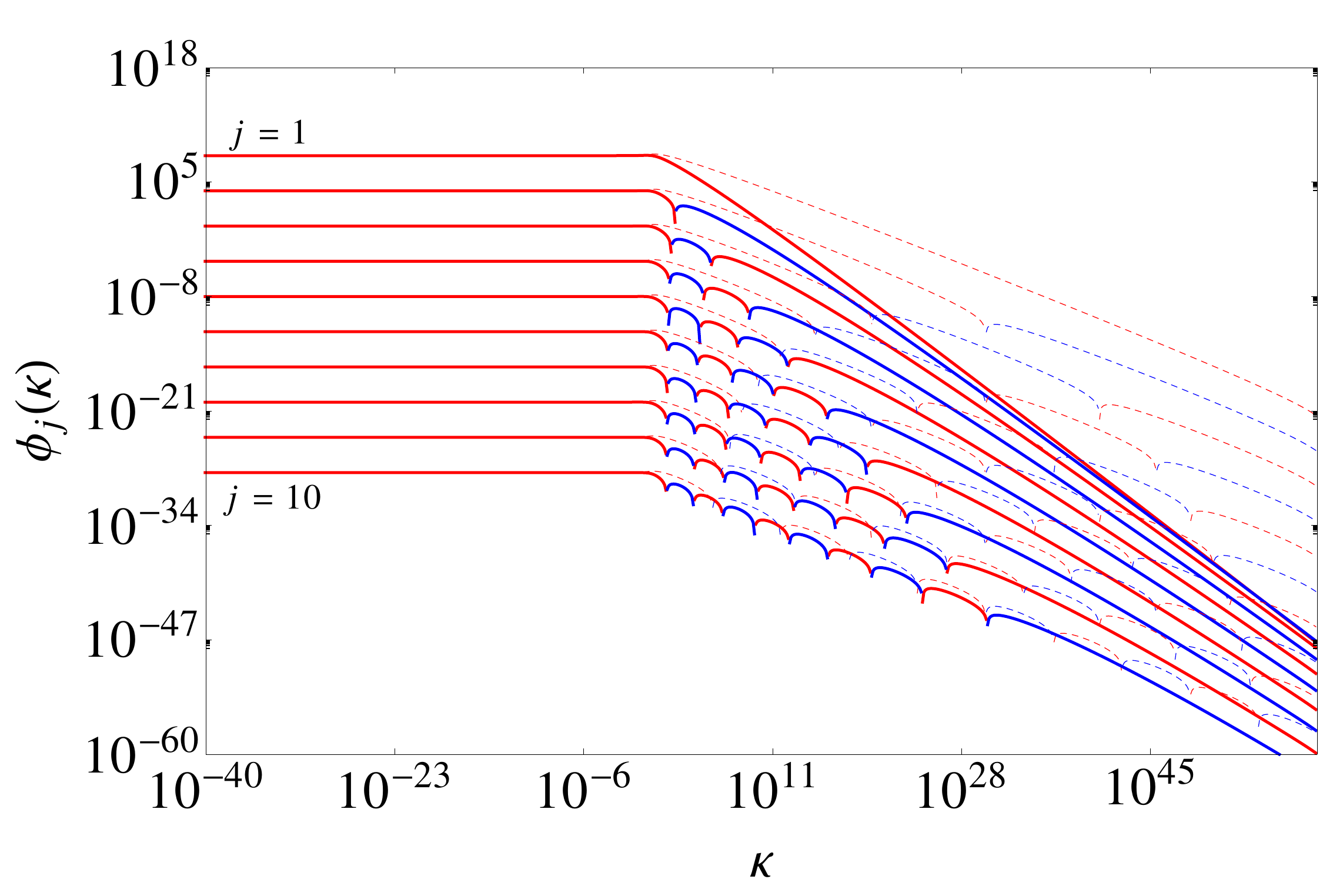}\protect\protect\protect\caption{\label{fig:R_WF}(color online) Wave functions with running coupling
which correspond to the first 10 eigenvalues (multiplied by arbitrary
constant factor for better visibility). As expected, after ordering
by eigenvalues, a number of nodes in the $j$th wave function equals
$j-1$. Changing color (red-blue) reflects change of sign from positive
to negative. Thin dashed lines show the results with fixed coupling
obtained in our earlier paper~\cite{LLS1}, normalization chosen
to match results with running coupling at small-$\kappa$.}
\end{figure}

Finally, in the Figure~\ref{fig:R_WF} we plot the wave functions
for the first ten eigenvalues and compare them with wave functions
evaluated with fixed coupling scheme. As we can see, the falloff at
large momenta is much faster than for the case of fixed coupling,
in agreement with~(\ref{GRA5}). Also, we note that the nodes are
no longer equidistant in the logarithmic coordinates.

We conclude that a running coupling~(\ref{ALHOUT}) leads to discrete
spectrum, however, due to a sensitivity to the choice of the confinement
model, the leading intercept is subject to a sizable uncertainty.

\subsection{Results with triumvirate coupling~(\ref{eq:alpha_QCD})}

\label{sub:triumvirate}In Section~\ref{sec:Theory} we discussed
that the choice of the coupling~(\ref{ALHOUT}) is not the only possibility,
and the so-called triumvirate parametrization~(\ref{eq:alpha_QCD})
has certain advantages. However, the evaluation of the spectra in
semiclassical approximation in this case is impossible, and for this
reason, up to now the spectrum of BFKL pomeron with this coupling
has never been studied. In this paper we address this problem, and
evaluate the eigenvalues of Eq.~(\ref{EQF-1}) using the lattice
approach described in Appendix~\ref{sec:NumericalMethod}.

In order to regularize the infrared pole in the running coupling $\alpha_{s}$,
we introduce a freeze-out scale $\lambda m_{H}^{2}$ into its argument,

\begin{equation}
\alpha_{s}\left(\left(\vec{k}-\vec{k}^{\,'}\right)^{2}\right)\to\alpha_{s}\left(\left(\vec{k}-\vec{k}^{\,'}\right)^{2}+\lambda\, m_{H}^{2}\right).\label{eq:alpha_triumvirate}
\end{equation}

\begin{figure}
\includegraphics[width=13cm]{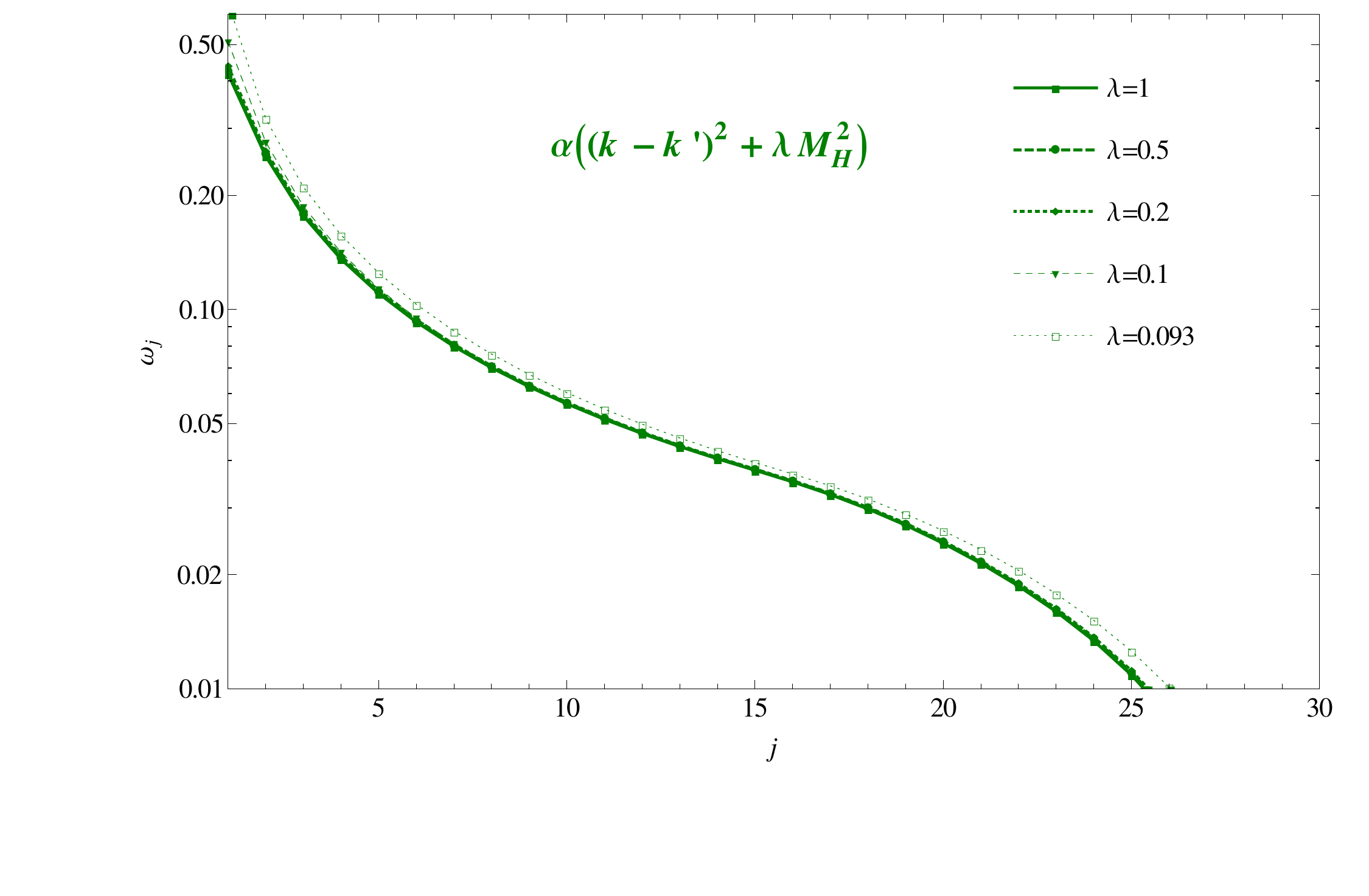}\\
 \includegraphics[width=13cm]{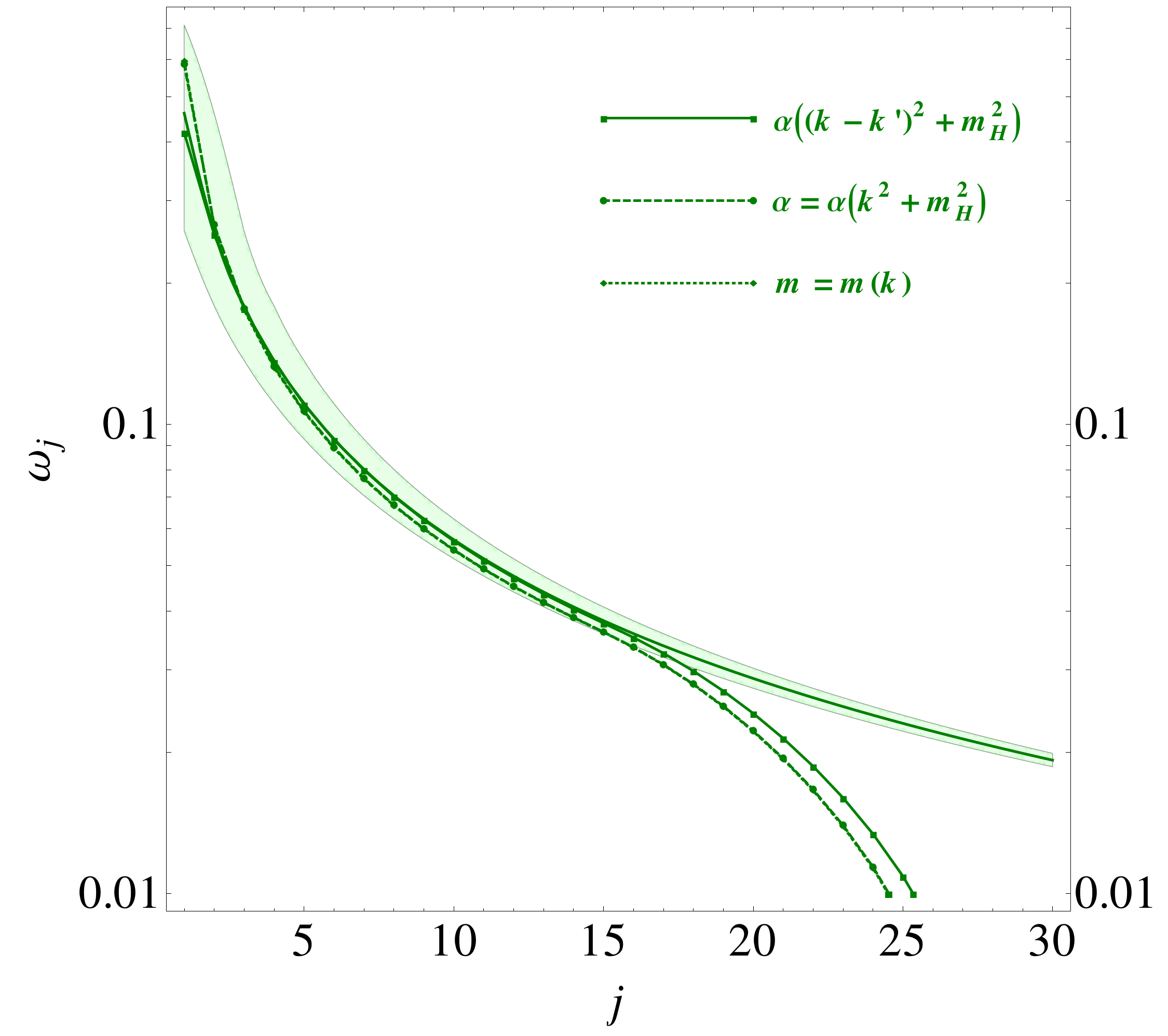}\protect\protect\caption{\label{fig:R3_IR}(color online) Upper plot: Eigenvalues of the BFKL
equation with a coupling~(\ref{eq:alpha_QCD}). A value of $\lambda$
was introduced to regularize infrared behavior of a coupling constant.
Lower plot: A comparison of the eigenvalues evaluated in different
IR regularization schemes: adding mass to the argument of a coupling
constant with coupling~(\ref{eq:alpha_QCD_simple}) (solid line)
and (\ref{eq:alpha_QCD}) (dashed line), as well as setting the same
infrared cutoff as the lowest momentum $k_{{\rm min}}$ (dotted line).
A value of $\lambda$ was introduced to regularize an infrared behavior
of the coupling constant. Light-green band corresponds to a result
of semiclassical approximation in~(\ref{SM21}), with band width
reflecting uncertainty due to nonperturbative phase~$\varphi^{\mbox{\tiny non-pert}}$;
a central line corresponds to a phase $\varphi^{\mbox{\tiny non-pert}}$
set as in~(\ref{SM4}). }
\end{figure}

We found that similar to the case of the coupling~(\ref{ALHOUT}),
the spectrum of the problem is discrete, which manifests itself in
significant lattice independent distances between neighboring eigenvalues.
From Fig.~\ref{fig:R3_IR} we can see that the sensitivity to the
choice of the infrared scale in~(\ref{eq:alpha_triumvirate}) is
quite mild, which is somewhat unexpected, given the fact that the
BFKL kernel~(first term in the rhs of~(\ref{EQF-1})) is strongly
peaked around $k\approx k'$. As we demonstrate in Appendix~\ref{sec:renormalon},
this happens because for large $k$ the contribution of this region
is suppressed as $\sim\Lambda_{{\rm QCD}}/k$ due to cancellations
of singularities of kernel $K$ at $k\approx k'$, and for this reason
there is a mild sensitivity to the choice of freeze-out scale. In
contrast to results of previous section, this sensitivity exists for
all eigenvalues (not only the leading intercept). However, numerically
eigenvalues of~(\ref{EQF-1}) and (\ref{EQF-2}) coincide (within
30\%) with each other. This coincidence is due to the above-mentioned
fact that the kernel~(\ref{EQF-1}) is peaked around $k\approx k'$,
and for this reason we expect that any other prescription for the
running coupling argument (like e.g.~\cite{Brodsky:1998kn,Brodsky:2002ka})
should lead to similar results.

\begin{figure}
\includegraphics[width=13cm]{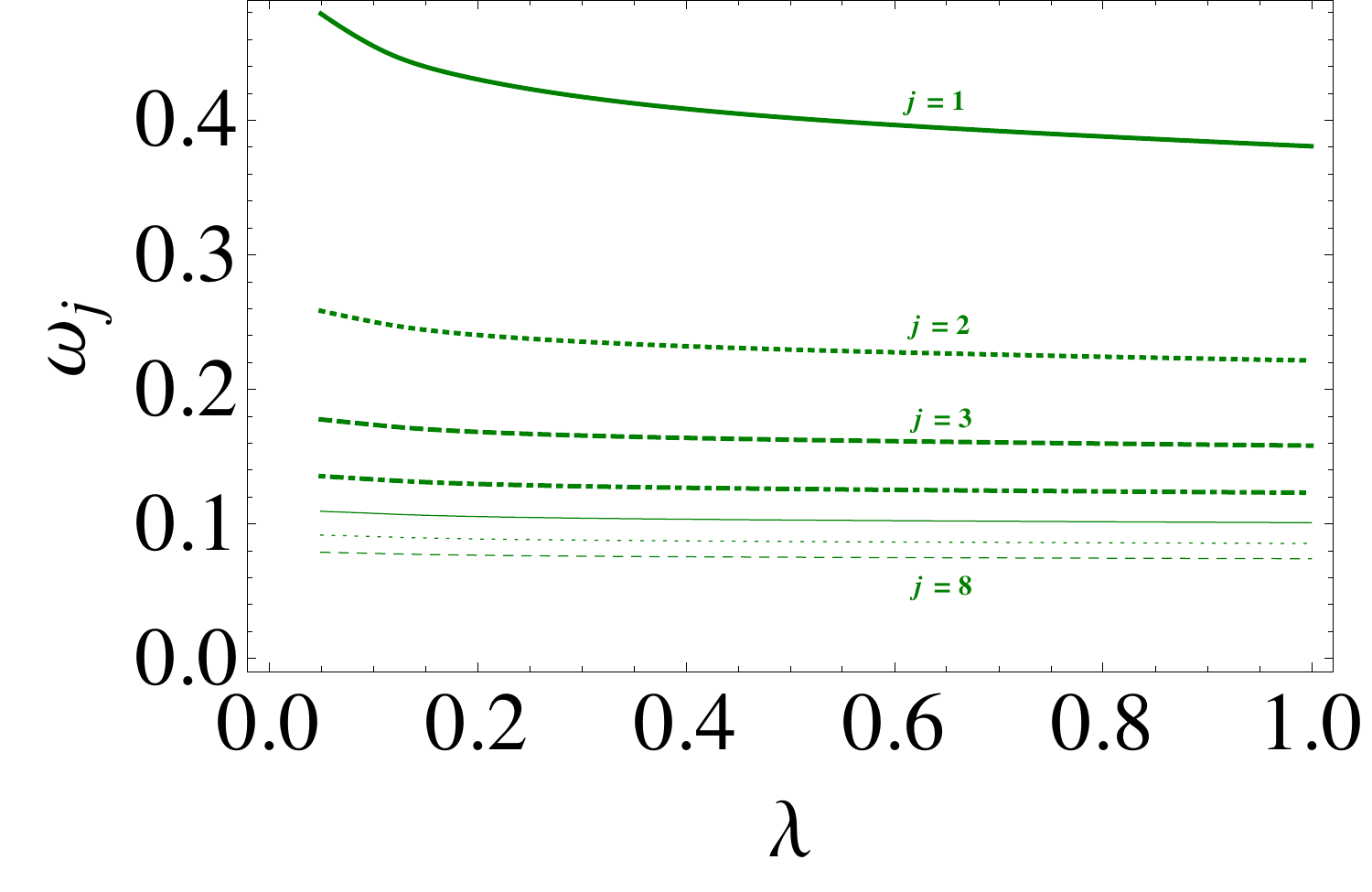}\protect\protect\protect\protect\caption{\label{fig:R3_IR-1}(color online) Dependence of several largest eigenvalues
of the BFKL equation with a coupling~(\ref{eq:alpha_eff_reg}) on
a value of infrared parameter $\lambda$. }
\end{figure}

Finally, in order to address the uncertainty in the infrared regularization,
we also consider another regularization scheme (see Appendix~\ref{sec:renormalon}
for details) 
\begin{equation}
\alpha_{{\rm eff}}\left(k,k'\right)\approx\frac{\bar{\alpha}_{S}(k)}{1\,+\,2\, b\,\bar{\alpha}_{S}(k)\,\ln\left(\frac{|k^{2}-k'^{2}|}{k^{2}}+\lambda\right)},\label{eq:alpha_eff_reg}
\end{equation}
where $\lambda$ is a small effective parameter. In the Table~(\ref{tab:TenRoots})
we give the first few eigenvalues for several values of $\lambda$.
As a function of eigenvalue number $j$, it has a behavior very similar
to~(\ref{eq:alpha_triumvirate}) (shown in Figure~\ref{fig:R3_IR}),
for this reason for the sake of legibility in the Figure~(\ref{fig:R3_IR-1})
we only show the dependence on $\lambda$. For $\lambda\approx1$
the corresponding eigenvalues are much smaller than with a scheme~(\ref{eq:alpha_triumvirate}),
since the former roughly corresponds to the latter with an artificially
increased infrared freezeout scale $\sim k^{2}$. However, in the
limit $\lambda\to0$, the difference between the two schemes vanishes.

\subsection{Lattice results \& continuum limit}

\label{sec:cutoff}In this paper we considered two different methods
to study the spectrum of BFKL, the semiclassical approximation and
the lattice method. The former due to limitations imposed by its applicability
is not very reliable for the first leading eigenvalues, with accuracy
improving considerably for larger eigenvalues. This is especially
important because these first roots determine the behavior of the
BFKL theory.

\begin{figure}
\includegraphics[width=9cm]{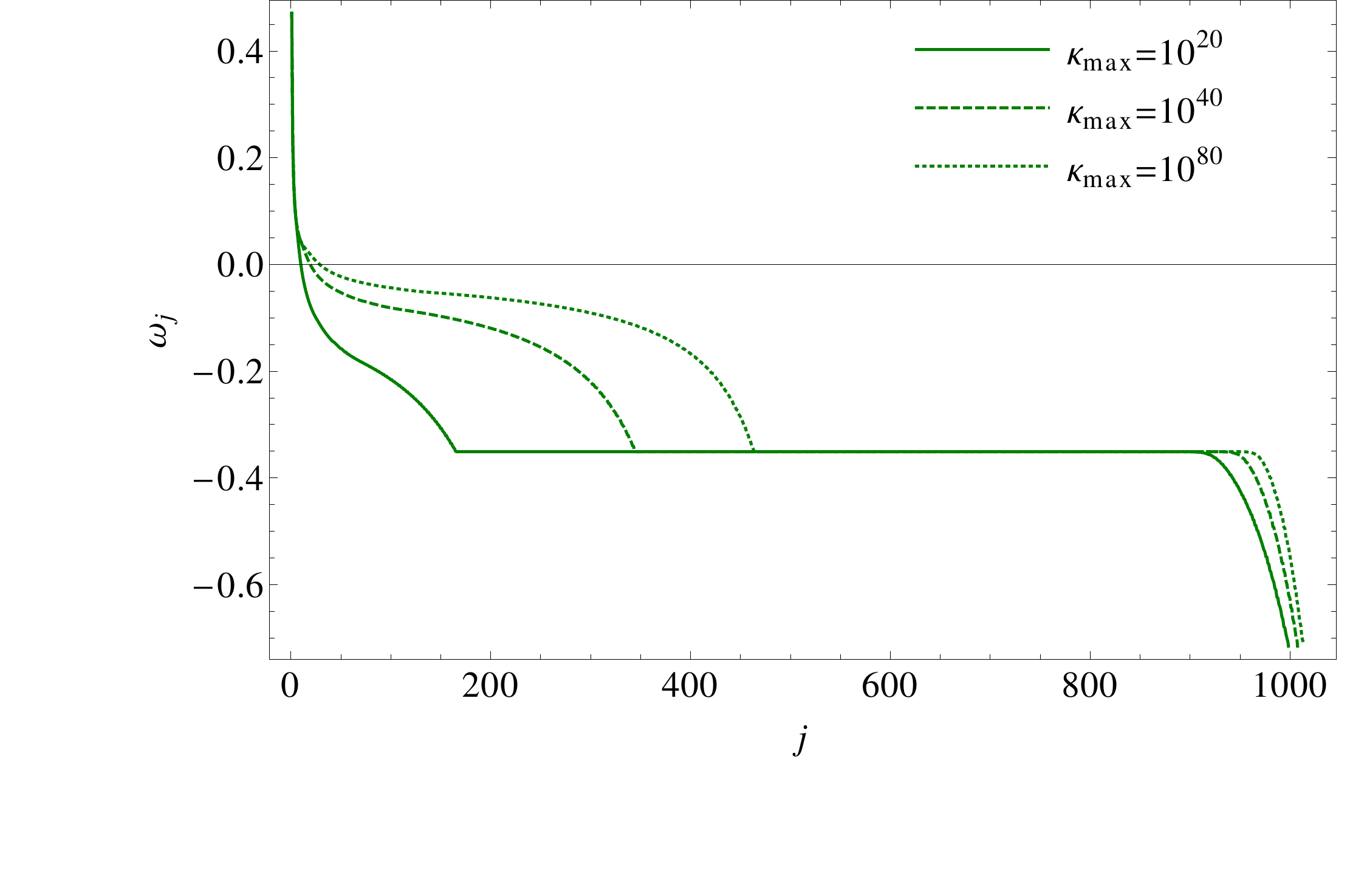}\includegraphics[width=9cm]{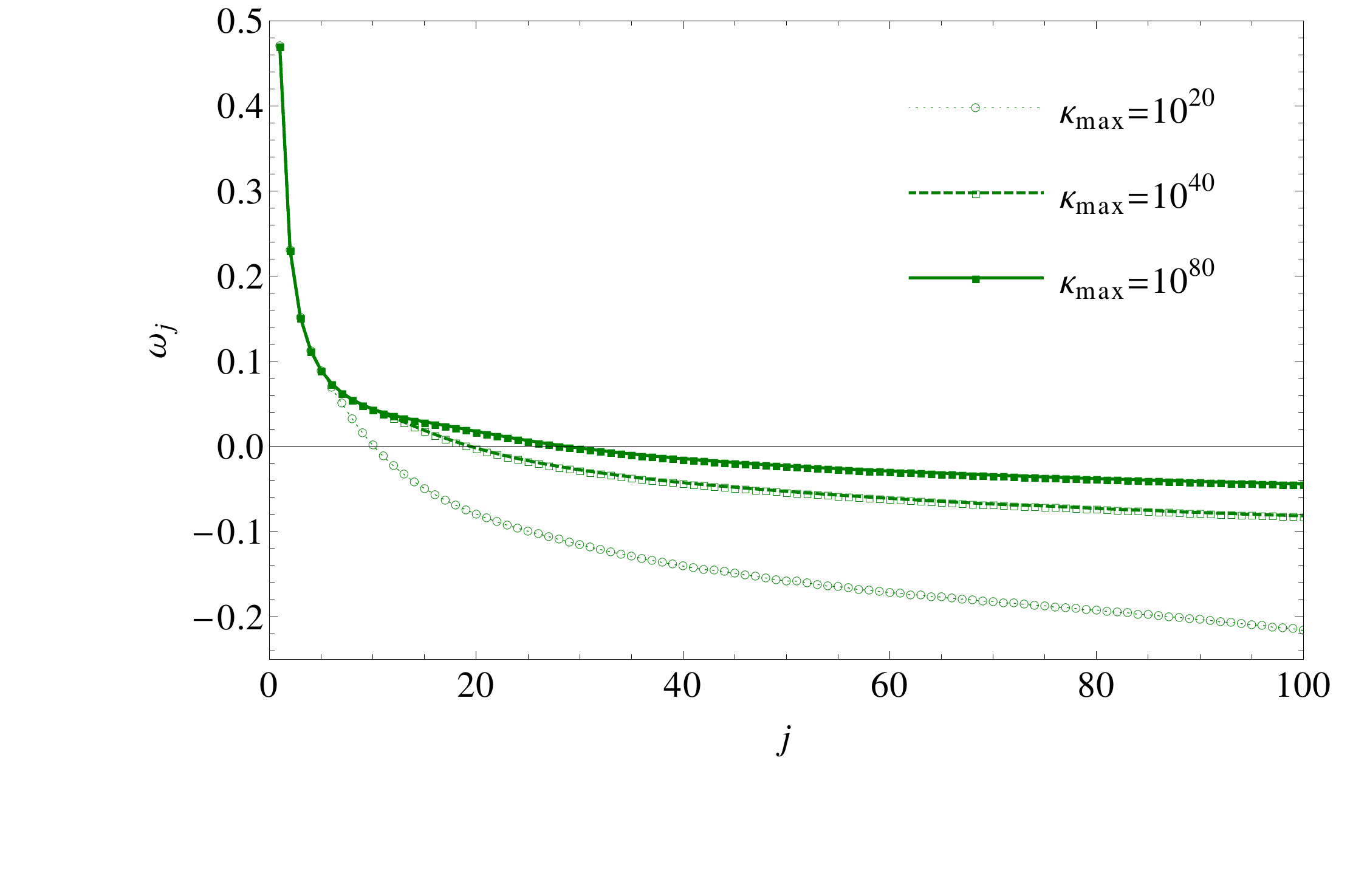}\protect\protect\caption{\label{cutoff} The eigenvalues of the BFKL equation with running
coupling~(\ref{ALHOUT}) versus the values of upper cutoff $\kappa_{max}$.
Right plot is a magnified version of the left plot for small values
of $j$.}
\end{figure}

On the contrary, the lattice is applicable for the leading intercepts,
but, as we can infer from Fig.~(\ref{fig:R_IR},\ref{fig:R3_IR}),
becomes unreliable for large root number $j\gg1$, when the size of
the lattice becomes insufficient to resolve the distance between neighboring
roots. As a consequence, instead of the infinite series~(\ref{SM4})
of discrete positive eigenvalues, the lattice sees only a finite number
of them, jumping to a negative branch (continuous spectrum). However,
as we can see from Figure~\ref{cutoff}, the number of positive eigenvalues
increases with increase of $\kappa_{max}$, and eventually reproduces
an infinite series~(\ref{SM4}). Similar to the case of fixed coupling~(\ref{EQF})
studied in~\cite{LLS1}, we observe that at negative $\omega$ there
is a plateau in $j$ -dependence: the spectrum freezes near a value
$\sim\bar{\alpha}_{s}\left(0\right)T\left(0\right)$, where $\bar{\alpha}_{s}\left(0\right)$
is a value at which running coupling freezes due to infrared regularization,
and $T(0)$ is a kinetic energy term (first term in~(\ref{EQF})).
Together, eigenfunctions corresponding to positive and negative eigenvalues
form a complete and orthogonal set of functions.

\section{Conclusions}

\label{sec:Conclusions} In this paper we applied lattice methods
for study of the BFKL spectrum with running QCD, and investigated
its dependence on the confinement model and change of running coupling
prescriptions. As a model of low-energy (confinement) region we used
the non-abelian theory with the Higgs mechanism of mass generation
for modeling the behavior of the wave function at large distances.
In Section~\ref{sub:LatticeResults} we demonstrated that a dependence
on the low energy model is weak for all eigenvalues except the leading
Regge pole. For the latter, we observed a sizable ($\sim$20-30\%)
sensitivity to details of confinement physics. We established that
semiclassical result for the leading intercept within uncertainty
of the phase%
\footnote{This phase uncertainty, which stems from the unknown confinement region,
is the dominant source of uncertainty in the semiclassical consideration
and is significantly larger than other sources of uncertainty discussed
in Section~(\ref{sec:QuasiClassics}). %
} coincides with exact lattice result, thus justifying application
of the semiclassical approach. Also, we estimated the leading eigenvalues
with variational method. We used the trial functions based on diffusion
approximation solutions of the BFKL equation and demonstrated that
this method yields leading eigenvalues consistent with the exact numerical
solutions.

In Section~\ref{sub:triumvirate}, for the first time we studied
the spectrum with the so-called triumvirate form of the running coupling~(\ref{eq:alpha_QCD})
and found that it is discrete. We found that a mild (up to 15\%) dependence
on the choice of the freezing scale exists for any eigenvalue, and
stems from $\mathcal{O}\left(\Lambda_{QCD}^{2}/k^{2}\right)$-contributions
of renormalons, as discussed in Appendix~\ref{sec:renormalon}. However,
numerically, eigenvalues of~(\ref{EQF-1}) and (\ref{EQF-2}) coincide
with each other within 10 percent..

\section*{Acknowledgments}

We thank our colleagues at Tel-Aviv and UTFSM university for encouraging
discussions. Our special thanks go to Asher Gotsman, whose comments
and suggestions were very instrumental for our work. L.L. acknowledges
partial support of the Russian Scientific Fund project 14-2200281.
This research was also partially supported by BSF grant 2012124, Proyecto
Basal FB 0821 (Chile), the Fondecyt (Chile) grants 1140842 and 1140377,
CONICYT (Chile) grants PIA ACT1406 and ACT1413. Powered@NLHPC: This
research was partially supported by the supercomputing infrastructure
of the NLHPC (ECM-02). Also, we thank Yuri Ivanov for technical support
of the USM HPC cluster where part of evaluations were done.

\appendix

\section{Study of eigenvalues in the lattice}

\label{sec:NumericalMethod}

The equations~(\ref{EQF-1},\ref{EQF-2},\ref{EQF}) have a common
form,

\begin{equation}
\omega\,\phi\left(\kappa\right)\,\,=\,\,\int\, d\kappa'\,\, K\left(\kappa,\kappa'\right)\,\phi\left(\kappa'\right)\label{NS1-1}
\end{equation}

where a kernel $K$ is a smooth function of both its arguments and
decreases at large momenta as $\sim1/{\rm max}(\kappa,\kappa')$ .
Such large-momenta asymptotics implies that in order to avoid sensitivity
to ultraviolet effects, a very large value of the upper cutoff $\kappa_{max}$
should be introduced. For the numerical calculation of eigenvalue,
we replace the continuous variables $\kappa$ and $\kappa'$ by the
discrete set of $\{\kappa_{n}\}$ and $\{\kappa'_{n}\}$ using the
logarithmic grid (in $\kappa=k^{2}/m^{2}$) with $N+1$ nodes,

\begin{eqnarray}
\kappa_{n} & = & \kappa_{min}\exp\left(\frac{n}{N}\,\ln\left(\kappa_{max}/\kappa_{min}\right)\right),\quad n=0,...,N,\label{NS2-1}
\end{eqnarray}

where the values of $\kappa_{min},\,\kappa_{max}$ were set to $\kappa_{min}=10^{-40},\,\kappa_{max}=10^{80}$,
and $N=1024$. The equation~(\ref{NS1-1}) turns into a discretized
linear problem.

As we demonstrated in Ref.\cite{LLS1}, for the fixed coupling constant
and massless limit, this method reproduces correctly the analytic
spectrum~(\ref{EIGV}), with a very mild sensitivity to further improvements
of lattice parameters. Due to finite size of the lattice, the spectrum
is discretized, with the distance between neighboring eigenvalues
$j$ and $j+1$ given by

\begin{equation}
\Delta\omega_{j+1,j}^{{\rm lattice}}=\bar{\alpha}\Big\{ b\,\frac{d\chi}{d\nu}\Big{|}_{\nu=b\, j}\,\,+\,\,\frac{1}{2}b^{2}(2j+1)\,\frac{d^{2}\chi}{d\nu^{2}}\Big{|}_{\nu=b\, j}\Big\}\label{eq:SpectrumSpacing}
\end{equation}

where parameter $b$ is defined in~(\ref{eq:b_def}) and vanishes
as $\sim1/\ln\kappa_{{\rm max}}$ in this limit. As we can get from
(\ref{eq:SpectrumSpacing}), this distance is of order 0.03-0.05 and
is considerably smaller than the distance between neighboring eigenvalues
for the first leading intercepts (\ref{EQF-1},\ref{EQF-2}), which
signals that the spectrum is discrete. However, for eigenvalues with
larger $j$ a lattice cannot reliably discern neighboring discrete
roots.

\section{The contribution of the infrared renormalon}

\label{sec:renormalon} In this section we discuss the contribution
of the infrared renormalon (IR) to the BFKL equation~(\ref{GRA1})
with the triumvirate coupling~(\ref{eq:alpha_QCD}) (see more in
Ref.\cite{LEAS,GKRW}). For the sake of simplicity we consider the
case of massless theory, assuming that the generalization for the
case of massive theory is straightforward. Explicitly, the equation~(\ref{GRA1})
can be rewritten in this case as 
\begin{equation}
\omega\phi_{\omega}\left(k\right)\,\,=\,\,\int d^{2}k'\,\frac{\bar{\alpha}_{S}\left(k'^{2}\right)}{\bar{\alpha}_{S}\left(k^{2}\right)}\,\frac{\bar{\alpha}_{S}\left(\left(\vec{k}\,-\,\vec{k}'\right)^{2}\right)}{\left(\vec{k}\,-\,\vec{k}'\right)^{2}}\,\Bigg\{\phi_{\omega}\left(k'\right)\,\,-\,\,\frac{k^{2}}{k'^{2}+\left(\vec{k}\,-\,\vec{k}'\right)^{2}}\,\phi_{\omega}\left(k\right)\Bigg\}.\label{BFKL}
\end{equation}
As we can see, a kinematic region $\left(\vec{k}\,-\,\vec{k}'\right)^{2}\leq\Lambda_{QCD}^{2}$
, $k\approx k'\gg\Lambda_{QCD}$ can potentially lead to a divergent
contribution and thus deserves special attention. In order to integrate
over the azimuthal angle $\varphi$, we rewrite a running coupling
as 
\begin{eqnarray}
\bar{\alpha}_{S}\left(\left(\vec{k}\,-\,\vec{k}'\right)^{2}\right)\, & = & \,\frac{\bar{\alpha}_{S}\left(k\right)}{1\,+\,\beta_{0}\,\bar{\alpha}_{S}\left(k\right)\ln\left(\left(\vec{k}-\vec{k}^{\,'}\right)^{2}/k^{2}\right)}\,=\,\bar{\alpha}_{S}\left(k^{2}\right)\,\sum_{i=0}^{\infty}\left(-\bar{\alpha}_{S}\left(k^{2}\right)\right)^{i}\,\ln^{i}\left(\frac{\left(\vec{k}-\vec{k}^{\,'}\right)^{2}}{k^{2}}\right).\label{BFKL01}
\end{eqnarray}
After integration of each term over the angle, we obtain~%
\footnote{Here we use 3.665(2) and 9.131 from~\cite{RY}%
} 
\begin{eqnarray}
\int d\varphi\,\frac{\bar{\alpha}_{S}\left(\left(\vec{k}\,-\,\vec{k}'\right)^{2}\right)}{\left(\vec{k}\,-\,\vec{k}'\right)^{2}}\, & = & \,\,\bar{\alpha}_{S}\left(k^{2}\right)\,\sum_{i=0}^{\infty}\left(-\bar{\alpha}_{S}\left(k^{2}\right)\right)^{i}\,\frac{d^{i}}{d\mu^{i}}|_{\mu=0}\int\frac{d\varphi\,\left(k^{2}\right)^{-\mu}}{\left(\left(\vec{k}-\vec{k}^{\,'}\right)^{2}\right)^{1-\mu}}\nonumber \\
\, & = & \,\pi\bar{\alpha}_{S}\left(k^{2}\right)\,\sum_{i=0}^{\infty}\left(-\bar{\alpha}_{S}\left(k^{2}\right)\right)^{i}\,\frac{d^{i}}{d\mu^{i}}|_{\mu=0}\frac{1}{|k^{2}-k'^{2}|}\,\left(\frac{|k^{2}-k'^{2}|}{k^{2}}\right)^{2\mu}\,{}_{2}F_{1}\left(\mu,\mu,1,z\right)\,\nonumber \\
 & \rightarrow & \,\frac{\pi}{\left|k^{2}-k'^{2}\right|}\left(\frac{\bar{\alpha}_{S}\left(k^{2}\right)}{1\,+\,2\,\beta_{0}\bar{\alpha}_{S}\left(k^{2}\right)\,\ln\left(\frac{|k^{2}-k'^{2}|}{k^{2}}\right)}\,+\,{\cal O}\left(z\bar{\alpha}_{S}^{3}\left(k^{2}\right)\right)\right),
\end{eqnarray}
where we introduced $z=k'^{2}/k^{2}$ for $k'<k$ and $z=k^{2}/k'^{2}$
for $k'>k$. Expanding the term in brackets $\{\dots\}$ in~(\ref{BFKL}),
we obtain 
\begin{eqnarray}
\omega\phi_{\omega}\left(k\right)\,\, & = & \,\,\int d^{2}k'\,\frac{\bar{\alpha}_{S}\left(k'^{2}\right)}{\bar{\alpha}_{S}\left(k^{2}\right)}\,\frac{\bar{\alpha}_{S}\left(\left(\vec{k}\,-\,\vec{k}'\right)^{2}\right)}{\left(\vec{k}\,-\,\vec{k}'\right)^{2}}\,\label{BFKL1}\\
 & \times & \,\,\Bigg\{\Big(1-\frac{k^{2}}{k'^{2}}\,+\,\frac{k^{2}\left(\vec{k}\,-\,\vec{k}'\right)^{2}}{k'^{4}}\Big)\phi_{\omega}\left(k\right)\,+\,\frac{k^{2}-k'^{2}}{k^{2}}\,\frac{d\phi_{\omega}\left(k'\right)}{d\ln\left(k'^{2}/k^{2}\right)}|_{k=k'}\,\,+\,\,\dots\,\Bigg\}\nonumber 
\end{eqnarray}

Let us first consider the first term in $\{\dots\}$ in which we put
$\bar{\alpha}_{S}\left(k^{2}\right)=\bar{\alpha}_{S}\left(k'^{2}\right)$
in our kinematic region, 
\begin{equation}
\int d^{2}k'\frac{\bar{\alpha}_{S}\left(\left(\vec{k}\,-\,\vec{k}'\right)^{2}\right)}{\left(\vec{k}\,-\,\vec{k}'\right)^{2}}\,\,\frac{k'^{2}\,-\, k^{2}}{k'^{2}}\xrightarrow{after\, angle\, integration}\int\frac{dk'^{2}}{k'^{2}}\frac{k'^{2}-k^{2}}{|k'^{2}-k^{2}|}\,\,\frac{\bar{\alpha}_{S}(k)}{1\,+\,2\, b\,\bar{\alpha}_{S}(k)\,\ln\left(\frac{|k^{2}-k'^{2}|}{k^{2}}\right)}.\label{BFKL2}
\end{equation}

\begin{enumerate}
\item In the region $k'>k$ we have 
\begin{equation}
\int\frac{dk'^{2}}{k'^{2}}\frac{k'^{2}-k^{2}}{|k'^{2}-k^{2}|}\,\,\frac{\bar{\alpha}_{S}(k)}{1\,+\,2\, b\,\bar{\alpha}_{S}(k)\,\ln\left(\frac{|k^{2}-k'^{2}|}{k^{2}}\right)}\,\,=\,\,\int_{1}^{\infty}dz\,\frac{\bar{\alpha}_{S}(k)}{1\,+\,2\, b\,\bar{\alpha}_{S}(k)\,\ln\left(z-1\right)}\label{BFKL3}
\end{equation}
where we introduced $z=k'^{2}/k^{2}$. Although the expansion~(\ref{BFKL1})
is valid only for $k'\approx k$, for a moment we will ignore this
fact and extend the upper integration limit to $\infty.$ Deriving
Eq.~(\ref{BFKL3}) we replaced $dk'^{2}/k'^{2}\rightarrow dk'^{2}/k^{2}$.
Introducing a new variable $z-1=e^{\frac{1}{2}u}$ we obtain 
\begin{equation}
\int_{1}^{\infty}dz\,\frac{\bar{\alpha}_{S}(k)}{1\,+\,2\, b\,\bar{\alpha}_{S}(k)\,\ln\left(z-1\right)}\,\,=\,\frac{1}{2}\,\int_{-\infty}^{+\infty}du\, e^{\frac{1}{2}u}\frac{\bar{\alpha}_{S}(k)}{1\,+\, b\,\bar{\alpha}_{S}(k)\, u}\label{BFKL4}
\end{equation}

\item In the region $k>k'$ we have 
\begin{equation}
\int\frac{dk'^{2}}{k'^{2}}\frac{k'^{2}-k^{2}}{|k'^{2}-k^{2}|}\,\,\frac{\bar{\alpha}_{S}(k)}{1\,+\,2\, b\,\bar{\alpha}_{S}(k)\,\ln\left(\frac{|k^{2}-k'^{2}|}{k^{2}}\right)}\,\,=\,\,-\,\int_{0}^{1}dz\,\frac{\bar{\alpha}_{S}(k)}{1\,+\,2\, b\,\bar{\alpha}_{S}(k)\,\ln\left(1-z\right)}\label{BFKL5}
\end{equation}

Introducing $1-z\,=\, e^{\frac{1}{2}u}$ we reduce Eq.~(\ref{BFKL5})
to 
\begin{equation}
\,\frac{1}{2}\,\int_{0}^{-\infty}du\, e^{\frac{1}{2}u}\,\frac{\bar{\alpha}_{S}(k)}{1\,+\,\, b\,\bar{\alpha}_{S}(k)\, u}\,=\,-\,\,\frac{1}{2}\,\int_{-\infty}^{0}du\,\, e^{\frac{1}{2}u}\frac{\bar{\alpha}_{S}(k)}{1\,+\,\, b\,\bar{\alpha}_{S}(k)\, u}\label{BFKL6}
\end{equation}

\end{enumerate}
Summing Eq.~(\ref{BFKL4}) and Eq.~(\ref{BFKL6}) we see that 
\begin{equation}
\int\frac{dk'^{2}}{k'^{2}}\frac{k'^{2}-k^{2}}{|k'^{2}-k^{2}|}\,\,\frac{\bar{\alpha}_{S}(k)}{1\,+\,2\, b\,\bar{\alpha}_{S}(k)\,\ln\left(\frac{|k^{2}-k'^{2}|}{k^{2}}\right)}\,\,=\,\,\frac{1}{2}\,\int_{0}^{+\infty}du\,\frac{e^{\frac{1}{2}u}}{1+e^{\frac{1}{2}u}}\,\,\frac{\bar{\alpha}_{S}(k)}{1\,+\, b\,\bar{\alpha}_{S}(k)\, u}\label{BFKL7}
\end{equation}
where we replaced $dk'^{2}/k^{2}\to dk'^{2}/k'^{2}$. Thus, in Eq.~(\ref{BFKL7})
the infrared renormalon completely cancels, and only ultraviolet renormalon
remains.

In all estimates in this section we neglect the contributions of the
order of $\Lambda_{QCD}^{2}/k^{2}$. We expect that such terms give
$\mathcal{O}\left(\Lambda_{QCD}^{2}/k^{2}\right)$-contribution from
the infrared renormalon which are negligible outside of confinement
region. The fact that the integral in~(\ref{BFKL7}) is only taken
over $u>0$ implies that the dominant contribution stems from region
$(k'^{2}-k^{2})/k^{2}>1$. The last term in the l.h.s. of~(\ref{BFKL7})
can be interpreted as an effective (angular averaged) coupling constant
\begin{equation}
\alpha_{{\rm eff}}\left(k,k'\right)\approx\frac{\bar{\alpha}_{S}(k)}{1\,+\,2\, b\,\bar{\alpha}_{S}(k)\,\ln\left(\frac{|k^{2}-k'^{2}|}{k^{2}}\right)},\label{eq:alpha_eff}
\end{equation}

which vanishes near the point $k\approx k'$. In order to regularize
a behavior of~(\ref{eq:alpha_eff}) near $k\approx k'$, we introduce
in the argument of logarithm a small constant $\lambda$, which leads
to an effective coupling~(\ref{eq:alpha_eff_reg}). This form of
regularization is inspired by the fact that a relevant parameter in
the expansion~(\ref{BFKL1}) is the ratio~$\left(k^{2}-k'^{2}\right)/k^{2}$
rather than $\left(k^{2}-k'^{2}\right)/\Lambda_{{\rm QCD}}^{2}$.
Besides similar parameter controls the size of higher order corrections
in expansion near $k'\approx k$: as can be seen from the structure
of the third term in~(\ref{BFKL1}), its relative size is given by
\begin{equation}
\mathcal{R}=\frac{k'^{2}-k^{2}}{k^{2}}\left.\frac{d\ln\phi_{\omega}\left(k'\right)}{d\ln k'^{2}}\right|_{k'=k}\approx\lambda\frac{d\ln\phi_{\omega}\left(k'\right)}{d\ln k'^{2}},
\end{equation}
where $\lambda$ is some parameter. As can be seen from~(\ref{APPWFCS}),
the ratio ~$d\ln\phi_{\omega}\left(k'\right)/d\ln k'^{2}\sim1$,
for this reason $\mathcal{R}\approx\lambda$. 


\begin{thebibliography}{10}
\bibitem{BFKL} E. A. Kuraev, L. N. Lipatov, and F. S. Fadin, \textit{Sov.
Phys. JETP} \textbf{45}, 199 (1977); \,\,\, Ya. Ya. Balitsky and
L. N. Lipatov, \textit{Sov. J. Nucl. Phys.}\, \textbf{28}, 22 (1978).

\bibitem{LI} L. N. Lipatov, Phys. Rep. \textbf{286} (1997) 131; Sov.
Phys. JETP \textbf{63} (1986) 904 {[}Zh.\ Eksp.\ Teor.\ Fiz.\ \textbf{90},
1536 (1986){]}.

\bibitem{KLB} Yuri V Kovchegov and Eugene Levin, \textit{``Quantum
Choromodynamics at High Energies}'', Cambridge Monographs on Particle
Physics, Nuclear Physics and Cosmology, Cambridge University Press,
2012 and references therein.

\bibitem{FROI} M.~Froissart, \textit{Phys.\, Rev.} \, \textbf{123}
(1961) 1053; \\
 ~A. ~Martin, ``Scattering Theory: Unitarity, Analitysity and Crossing.''
Lecture Notes in Physics, Springer-Verlag, Berlin-Heidelberg-New-York,
1969.

\bibitem{KW1} A.~Kovner and U.~A.~Wiedemann, Phys.\ Rev.\ D
\textbf{66}, 051502 (2002) {[}hep-ph/0112140{]}.

\bibitem{KW2} A.~Kovner and U.~A.~Wiedemann, Phys.\ Rev.\ D
\textbf{66}, 034031 (2002) {[}hep-ph/0204277{]}.

\bibitem{KW3} A.~Kovner and U.~A.~Wiedemann, Phys.\ Lett.\ B
\textbf{551}, 311 (2003) {[}hep-ph/0207335{]}.

\bibitem{FIIM} E.~Ferreiro, E.~Iancu, K.~Itakura and L.~McLerran,
Nucl.\ Phys.\ A \textbf{710}, 373 (2002) {[}hep-ph/0206241{]}.

\bibitem{GLR} L. V. Gribov, E. M. Levin and M. G. Ryskin, Phys. Rep.
\textbf{100} (1983) 1.

\bibitem{LEAS} E.~Levin, Nucl.\ Phys.\ B \textbf{453} (1995) 303
{[}hep-ph/9412345{]}.

\bibitem{LERUN} E.~Levin, Nucl.\ Phys.\ B \textbf{545}, 481 (1999),
hep-ph/9806228.

\bibitem{KLR1} H.~Kowalski, L.~N.~Lipatov and D.~A.~Ross, Eur.\ Phys.\ J.\ C
\textbf{76} (2016) 1, 23, {[}arXiv:1508.05744 {[}hep-ph{]}{]}.

\bibitem{KLR2} H.~Kowalski, L.~Lipatov and D.~Ross, Eur.\ Phys.\ J.\ C
\textbf{74} (2014) 6, 2919, {[}arXiv:1401.6298 {[}hep-ph{]}{]}.

\bibitem{KLR3} H.~Kowalski, L.~N.~Lipatov and D.~A.~Ross, Phys.\ Part.\ Nucl.\ \textbf{44}
(2013) 547, arXiv:1205.6713 {[}hep-ph{]}.

\bibitem{KLR4} H.~Kowalski, L.~N.~Lipatov and D.~A.~Ross, \textit{``Indirect
Evidence for New Physics at the 10 TeV Scale,''} arXiv:1109.0432
{[}hep-ph{]}.

\bibitem{KLRW1} H.~Kowalski, L.~N.~Lipatov, D.~A.~Ross and G.~Watt,
Nucl.\ Phys.\ A \textbf{854}, 45 (2011);\,\,\, Eur.\ Phys.\ J.\ C
\textbf{70}, 983 (2010) {[}arXiv:1005.0355 {[}hep-ph{]}{]}.

\bibitem{KLRW2} H.~Kowalski, L.~N.~Lipatov, D.~A.~Ross and G.~Watt,
Eur.\ Phys.\ J.\ C \textbf{70} (2010) 983, {[}arXiv:1005.0355 {[}hep-ph{]}{]}.

\bibitem{LEPO} E.~Levin and I.~Potashnikova, JHEP \textbf{1402}
(2014) 089, {[}arXiv:1307.7823 {[}hep-ph{]}{]}.

\bibitem{LLS1} E.~Levin, L.~Lipatov and M.~Siddikov, Phys.\ Rev.\ D
\textbf{89} (2014) 7, 074002, {[}arXiv:1401.4671 {[}hep-ph{]}{]}.

\bibitem{LLS2} E.~Levin, L.~Lipatov and M.~Siddikov, Eur.\ Phys.\ J.\ C
\textbf{75} (2015) 11, 558, {[}arXiv:1508.04118 {[}hep-ph{]}{]}.

\bibitem{GRCO} V.~N.~Gribov, Nucl.\ Phys.\ B \textbf{139}, 1
(1978).

\bibitem{GRCTH} J.~Serreau, M.~Tissier and A.~Tresmontant, Phys.\ Rev.\ D
\textbf{89}, 125019 (2014) {[}arXiv:1307.6019 {[}hep-th{]}{]}\,\,\,
J.~Serreau and M.~Tissier, Phys.\ Lett.\ B \textbf{712}, 97 (2012)
{[}arXiv:1202.3432 {[}hep-th{]}{]}.

\bibitem{GRCREV} N.~Vandersickel and D.~Zwanziger, Phys.\ Rept.\ \textbf{520},
175 (2012) {[}arXiv:1202.1491 {[}hep-th{]}{]} and references therein.

\bibitem{GRCMASSG} J.~M.~Cornwall, Mod.\ Phys.\ Lett.\ A \textbf{28},
1330035 (2013) {[}arXiv:1310.7897 {[}hep-ph{]}{]}; J.~A.~Gracey,
J.\ Phys.\ A \textbf{47} (2014) 44, 445401, {[}arXiv:1409.0455 {[}hep-ph{]}{]}.

\bibitem{GRCLAT} P.~J.~Silva, D.~Dudal and O.~Oliveira, \textit{``Spectral
densities from the lattice,''} PoS LATTICE \textbf{2013} (2014) 366
{[}arXiv:1311.3643 {[}hep-lat{]}{]}.


P.~J.~Silva, O.~Oliveira, D.~Dudal, P.~Bicudo and N.~Cardoso,
\textit{b``Many faces of the Landau gauge gluon propagator at zero
and finite temperature: positivity violation, spectral density and
mass scales,''} PoS QCD \textbf{-TNT-III}, 040 (2013) {[}arXiv:1401.1554
{[}hep-lat{]}{]}.

\bibitem{LICRO} L.~N.~Lipatov, \textit{``Non-perturbative effects
for the BFKL equation in QCD and in $N=4$ SUSY,''} AIP Conf.\ Proc.\ \textbf{1654}
(2015) 070004.

\bibitem{BRAS} M.~A.~Braun, Phys.\ Lett.\ B \textbf{348} (1995)
190, {[}hep-ph/9408261{]}.

\bibitem{KOAS} Y.~V.~Kovchegov and H.~Weigert, Nucl.\ Phys.\ A
\textbf{784} (2007) 188 {[}hep-ph/0609090{]}.

\bibitem{BAAS} I.~Balitsky, Phys.\ Rev.\ D \textbf{75} (2007)
014001, {[}hep-ph/0609105{]}.

\bibitem{Brodsky:1998kn}S.~J.~Brodsky, V.~S.~Fadin, V.~T.~Kim,
L.~N.~Lipatov and G.~B.~Pivovarov, JETP Lett.~\textbf{70}, 155
(1999) {[}hep-ph/9901229{]}.

\bibitem{Brodsky:2002ka}S.~J.~Brodsky, V.~S.~Fadin, V.~T.~Kim,
L.~N.~Lipatov and G.~B.~Pivovarov, JETP Lett.~\textbf{76}, 249
(2002) {[}Pisma Zh.~Eksp.~Teor.~Fiz.~\textbf{76}, 306 (2002){]}
{[}hep-ph/0207297{]}.

\bibitem{DeRujula:1976edq}A.~De Rujula and H.~Georgi, Phys.~Rev.~D
\textbf{13}, 1296 (1976).

\bibitem{FADLI} V.~S.~Fadin and L.~N.~Lipatov, Phys.\ Lett.\ B
\textbf{429}, 127 (1998) {[}hep-ph/9802290{]};\,\,\, M.~Ciafaloni
and G.~Camici, Phys.\ Lett.\ B \textbf{430}, 349 (1998) {[}hep-ph/9803389{]}.

\bibitem{KOAS1} Y.~V.~Kovchegov and H.~Weigert, 
Nucl.\ Phys.\ A \textbf{789}, 260 (2007) {[}hep-ph/0612071{]}. 

\bibitem{DGLAP} V. N. Gribov and L. N. Lipatov, \textit{Sov. J. Nucl.
Phys} \textbf{15} (1972) 438;\\
 G. Altarelli and G. Parisi, \textit{Nucl. Phys.} \textbf{B 126} (1977)
298; \\
 Yu. l. Dokshitser, \textit{Sov. Phys. JETP} \textbf{46} (1977) 641.

\bibitem{Cornwall:1981zr}J.~M.~Cornwall, Phys.~Rev.~D \textbf{26},
1453 (1982).

\bibitem{Maris:2003vk}P.~Maris and C.~D.~Roberts, Int.~J.~Mod.~Phys.~E
\textbf{12}, 297 (2003)

\bibitem{Bartels:2006kr}J.~Bartels, L.~N.~Lipatov and K.~Peters,
Nucl.~Phys.~B \textbf{772}, 103 (2007) {[}hep-ph/0610303{]}.

\bibitem{Ross:2016zwl}D.~A.~Ross and A.~S.~Vera, arXiv:1605.08265
{[}hep-ph{]}.

\bibitem{GKRW} E.~Gardi, J.~Kuokkanen, K.~Rummukainen and H.~Weigert,
Nucl.\ Phys.\ A \textbf{784} (2007) 282, {[}hep-ph/0609087{]}. 

\bibitem{RY} I. Gradstein and I. Ryzhik, \textit{Table of Integrals,
Series, and Products}, Fifth Edition, Academic Press, London, 1994.


{[}hep-ph/9806482{]}.

\end{thebibliography}
\end{document}